\documentclass[twocolumn]{raa_twocolumn}           

\usepackage{graphicx,times}
\usepackage{natbib}
\usepackage{amssymb,amsmath}
\usepackage{comment}
\bibpunct{(}{)}{;}{a}{}{,}

\usepackage[utf8]{inputenc}

\usepackage[pagebackref=false]{hyperref}
\hypersetup{
    colorlinks=true,      
    linkcolor=black,      
    citecolor=blue,       
    urlcolor=blue         
}
\usepackage{cuted}
\usepackage{makecell}
\usepackage{multicol}
\usepackage{subcaption}
\usepackage[T1]{fontenc}

\begin{document}

   \title{A comprehensive multi-wavelength study of two X-ray emitting Be stars}

 \volnopage{ {\bf 20XX} Vol.\ {\bf X} No. {\bf XX}, 000--000}
   \setcounter{page}{1}

   \author{Hema Anilkumar \inst{1, 3}, Blesson Mathew \inst{1, 3}, Savithri H. Ezhikode \inst{2}, Sreeja S. Kartha \inst{1, 3}, Suman Bhattacharyya \inst{1, 3}, Sneha Nedhath \inst{4, 5}, Ajith N \inst{6}
   }

   \institute{Department of Physics and Electronics, CHRIST (Deemed to be University), Bangalore 560029, 
India; {\it hema.anilkumar@res.christuniversity.in, blesson.mathew@christuniversity.in}\\
        \and
            St. Francis de Sales College (Autonomous), Electronics City, Bengaluru - 560100, India
        \and    
            The Centre of Excellence in Astronomy and Astrophysics (CEAA), CHRIST (Deemed to be University), Bengaluru, India
        \and
            INAF -- Osservatorio Astrofisico di Arcetri, Largo E. Fermi 5, I-50125 Firenze, Italy
        \and
            Dipartimento di Fisica e Astronomia, Università di Firenze, Via G. Sansone 1,I-50019 Sesto Fiorentino (Firenze), Italy
        \and
            Department of Science, Christ Academy Institute for Advanced Studies, Bangalore 560083, India\\
\vs \no
   {\small Received 20XX Month Day; accepted 20XX Month Day}
}

\abstract{We present a multi-wavelength study of two X-ray-emitting Be stars, CXOU\,J052218.4+332820 (Star\,1) and CXOU\,J052304.2+332846 (Star\,2), identified from the \textit{LAMOST\,DR5 Hot emission-line stars catalog}. Our spectroscopic and photometric analyses indicate that Star\,1 is a main-sequence B5e star, whereas Star\,2 is a pre-main-sequence B7e star. The LAMOST spectrum of Star\,1 exhibits shell-type Fe\textsc{ii}, Paschen, and O\textsc{i} emission from circumstellar regions extending to $\sim$\,3--5\,R$_{\star}$. \textit{TESS} period analysis reveals a rapid rotational period of $\sim$\,0.355\,d and an inclination angle of $\sim$\,68\textdegree, confirming the Classical Be nature of Star\,1. In contrast, Star\,2 shows strong accretion signatures, including inverse P-Cygni Fe\textsc{ii} and redshifted He\textsc{i} absorption features. Its ZTF light curves display irregular dimming events of up to $\sim$\,3\,mag and quasi-periodic variability on timescales of $\sim$\,50--110\,d, consistent with circumstellar obscuration in an accreting Herbig Be star. To further investigate the high-energy properties of the two systems, we analyzed their \textit{Chandra} X-ray spectra. \textit{Chandra} spectroscopy yields moderate plasma temperatures (kT\,$\sim$\,1.7--1.8\,keV) and X-ray luminosities of logL$_{\rm X}$\,$\sim$\,30.4--30.6\,erg\,s$^{-1}$, comparable to those observed in magnetically active coronae. Furthermore, the positional agreement between the \textit{Chandra} and Gaia coordinates indicates that the X-ray emission is spatially associated with the optical systems rather than a nearby resolved companion. These results suggest that the observed X-ray emission originates from magnetically influenced circumstellar activity, consistent with magnetically confined wind-shock (MCWS) or magnetically torqued disk scenarios.
\keywords{techniques: photometric – techniques: spectroscopic – techniques: imaging spectroscopy – stars: emission-line, Be – stars: pre-main-sequence – stars: massive – stars: coronae – X-rays: stars}
}

    \authorrunning{Anilkumar et al. }            
    \titlerunning{Characterization of X-rays emitting ELS}  
    \maketitle

%
\section{Introduction} \label{sec:introduction} 
Emission-line stars (ELS) of B-type (Be stars) are a class of stars exhibiting one or more Balmer emission lines in their optical spectra. Spanning different evolutionary phases, they include the pre-main-sequence (PMS) Herbig Be stars \citep[HBe;][]{1998ARA&A..36..233Waters,2023MNRAS.524.5166Nidhi}, non-supergiant Classical Be stars \citep[CBe;][]{2013A&ARv..21...69Rivinius,2021MNRAS.500.3926Banerjee}, and Be supergiants. In the present work, we focus on the former two categories, as the mechanism responsible for Balmer emission in Be supergiants is different. Although both CBe and HBe stars exhibit Balmer emission associated with material surrounding the star, the physical origin of this material differs. In HBe stars, the emission lines observed in their spectrum originate from accreting disks and associated accretion flows linked to ongoing formation processes \citep{1998ARA&A..36..233Waters,Arun_2019}, while in CBe stars, they originate from a Keplerian decretion disk formed by material expelled due to rapid stellar rotation \citep{1982A&A...107..240Dachs,1994IAUS..162..399Waters,2008MNRAS.388.1879Mathew}. Both CBe and HBe stars exhibit significant photometric and spectroscopic variability. In CBe stars, variability is attributed to non-radial pulsations \citep[NRP;][]{2003A&A...411..229Rivinius, 2013A&ARv..21...69Rivinius}, disk-related changes, or rotational modulation \citep{1990MNRAS.245...92Balona, 2020MNRAS.493.2528Balona}. In contrast, variability in HBe stars is often linked to accretion-related phenomena and variable obscuration by dusty circumstellar material, leading to UX\,Ori-type behavior in some stars \citep[e.g.,][]{1998ARA&A..36..233Waters, 1999AJ....118.1043Herbst}. The variability studies provide key insights into the circumstellar environments and evolutionary processes of these stars.

Massive stars, particularly Be stars, are also observed to emit X-rays. In most cases, X-ray emission arises from shocks in their unstable, radiatively driven stellar winds, producing relatively soft X-rays (kT\,$\sim$\,0.2\,--\,0.6\,keV) with luminosities of L$_\mathrm{X}$/L$_\mathrm{bol}$\,$\sim$\,10$^{-7}$ \citep{Pallav1981, 1997A&A...322..167Berghoefer}. In HBe stars, the X-ray emission is attributed to magnetically heated coronae \citep[kT\,$\leq$\,2.0\,keV; L$_\mathrm{X}$\,$\leq$\,10$^{31}$\,erg\,s$^{-1}$; ][]{Stelzer2006, 2024MNRAS.530.3020Anilkumar}. In contrast, CBe stars are typically non-magnetic and are generally X-ray faint. However, weak surface magnetic fields of less than $\sim$150\,G have been detected in some cases \citep{2009IAUS..259..397Yudin,2009A&A...502..283Hubrig,2011IAUS..272..222Yudin}. A notable subclass known as $\gamma$ Cas analogs displays unusually hard (kT\,>\,5 keV) and luminous X-ray emission (L$_\mathrm{X}$\,>\,10$^\mathrm{32}$\,erg\,s$^{-1}$), which has been suggested to originate from processes such as magnetic interactions between the Be star and its decretion disk \citep{2016AdSpR..58..782Smith,2017A&A...602L...5Naze} or accretion onto a magnetic white dwarf companion \citep{2025A&A...703A.188Rauw}. The origin of X-ray emission in both CBe and HBe stars remains uncertain, with many studies exploring the role of magnetic fields in understanding the emission mechanism.

In this work, we determine the nature of the two X-ray emitting stars CXOU\,J052218.4+332820 (hereafter, Star\,1) and CXOU\,J052304.2+332846 (hereafter, Star\,2) from a multi-wavelength perspective. Though these stars have been catalogued earlier, they were not studied in detail. The paper is structured as follows: In section \ref{sec:Sample}, we discuss the selection procedure employed for the program stars. Section \ref{sec:data_reduction} provides details regarding the archival observations and data reduction. In section \ref{sec:analysisandresults}, we characterize the program stars discussing their spectral and photometric time-series properties individually. Further, in Section \ref{sec:Xrayproperties} we discuss the X-ray properties of these stars based on the parameters obtained from the temporal and spectral analysis of the \textit{Chandra} data. Finally, in Section \ref{sec:conclusion}, we summarize the results from our study.

\section{The Program Stars}\label{sec:Sample}
\begin{figure}[!t]
    \centering
    \includegraphics[width=0.9\columnwidth]{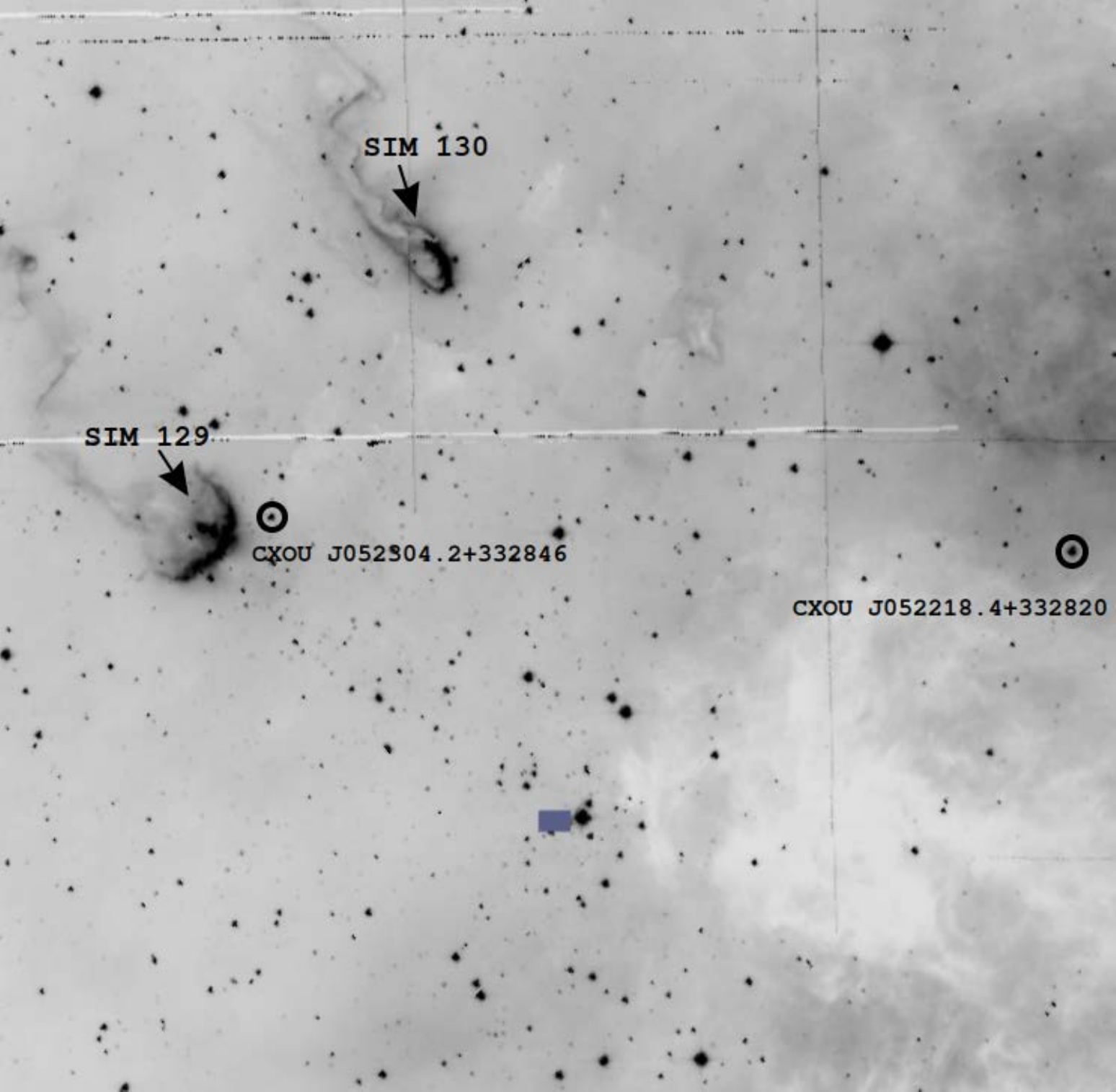}
    \caption{15\,arcsec\,$\times\,$15\,arcsec field of the cluster NGC\,1893 in the IPHAS H$\alpha$ narrow band filter. The targets CXOU\,J052218.4+332820 (Star\,1), and CXOU\,J052304.2+332846 (Star\,2) are marked by black circles. The two structures marked with arrows are SIM\,129 and SIM\,130, called the tadpoles or pennant nebula.} 
   \label{fig:clusterNGC1893}
\end{figure}

\cite{2021RAA....21..288Shridharan} studied a large sample of 3339 massive ELSs of spectral type O, B, and A, and categorized them into different stellar classes. We cross-matched this catalog with a list of all observations from the \textit{Chandra} data archive, and found nine ELSs with X-ray emission (detection rate $\sim$\,0.3\%). However, only two of these nine sources had good-quality optical spectra in the LAMOST\,DR8 Low-Resolution Spectroscopic (LRS) Survey. The remaining seven were excluded due to low signal-to-noise ratios (SNR), as our aim was to characterize these sources and understand the origin of their X-ray emission. Interestingly, the two ELSs (CXOU\,J052218.4+332820, and CXOU\,J052304.2+332846) are located within the field of the young open cluster NGC\,1893 \citep[$\sim$\,1\,--\,5\,Myr;][]{2007MNRAS.380.1141Sharma, 2013NewA...19....1Pandey}, which is situated within the H\textsc{ii} region IC\,410, at a distance of $\sim$\,3\,--\,6\,kpc \citep{2002A&A...393..195Marco, 2007MNRAS.380.1141Sharma}. The cluster lies toward the Galactic anti-center and hosts a rich population of early-type young stellar objects \citep[YSOs;][]{2007MNRAS.380.1141Sharma, 2011A&A...527A..77Prisinzano}. The cluster also features the two tadpole nebulae SIM\,129 and SIM\,130, commonly referred to as the pennant nebula (Figure \ref{fig:clusterNGC1893}).

\section{Observations and Data reduction}\label{sec:data_reduction}

\subsection{X-ray: \textit{Chandra} observations and data reduction}\label{subsec:chandradata}

We used the archival observations of the two program stars Star\,1 and Star\,2, which were carried out using the Advanced CCD Imaging Spectrometer Imaging array (ACIS-I) within the \textit{Chandra} X-ray observatory (CXO), during various epochs from 2006--11--07 to 2007--01--23. The observation details are tabulated in the Table \ref{tab:xraydat}. 

\begin{table}[!ht]
\centering
\caption{X-ray observation details from \textit{Chandra} X-ray observatory, covering the FOV of NGC\,1893, which include the two program stars: CXOU\,J052218.4+332820 (Star\,1) and CXOU\,J052304.2+332846 (Star\,2).}

\small
\setlength{\tabcolsep}{0.5pt}
\renewcommand{\arraystretch}{1.2}

\begin{tabular}{cccccc}
\hline 
\textbf{Region} & \textbf{RA [J2000]} & \textbf{Dec [J2000]} &
\textbf{Obs ID} & \textbf{Exp [ks]} & \textbf{Obs Date} \\
\hline \hline

NGC\,1893 & 80.70833 & 33.46805 & 08462 & 42.62 & 2006-11-07 \\
          &          &          & 06406 & 115.66 & 2006-11-09 \\
          &          &          & 06407 & 126.22 & 2006-11-15 \\
          &          &          & 08476 & 53.27 & 2006-11-17 \\
          &          &          & 06408 & 102.83 & 2007-01-23 \\
\hline
\end{tabular}

\label{tab:xraydat}
\end{table}

The \textit{Chandra} data was reduced utilizing the Chandra Interactive Analysis of Observations \citep[CIAO;][]{2006Fruscione} software package, version 4.16, along with the calibration database (CALDB)\footnote{\url{https://cxc.cfa.harvard.edu/caldb/downloads/index.html}} version 4.9.4. The standard data analysis procedures recommended by the \textit{Chandra X-ray Center (CXC)\footnote{\url{https://cxc.cfa.harvard.edu/ciao/guides/}}} were followed. The X-ray data analysis methodology adopted in this work follows the procedure described in detail in \cite{2024MNRAS.530.3020Anilkumar}. A brief summary is provided here for completeness. The level-2 raw event files obtained from the \textit{Chandra} data archive (ChaSeR) \footnote{\url{https://cda.harvard.edu/chaser/}} were reprocessed to produce a cleaned event file containing only the good time intervals (GTIs). An energy filter of 0.3\,--\,8.0 keV was applied to the cleaned event files. Further, we checked for background flaring with no significant flaring events identified. For source detection, we used the \texttt{wavdetect} algorithm with wavelet scales set to 1, 2, 4, and 8. A detection significance of 10$^{-6}$ is set to avoid false detections. A circular region for the source was set to the radius obtained from the \texttt{wavdetect} algorithm ($\sim$\,5\,arcsec). For the background, we chose a source-free region of radius $\sim$\,20\,arcsec within the same CCD. The X-ray light curves and spectra are then extracted. We discuss this in detail in section \ref{sec:Xrayproperties}.

\subsection{Optical: LAMOST\,DR8\,LRS observation details }\label{sec:opticalspectrometry}
We retrieved the optical spectra of our program stars from the Large Sky Area Multi-Object Fiber Spectroscopic Telescope (LAMOST)\,DR8\,v2.0 database\footnote{\url{https://www.lamost.org/dr8/v2.0/}}. The archival spectra of Star\,1 and Star\,2 were originally acquired on 2012--12--04 and 2015--01--20, respectively. Both the spectra are standard 1D pipeline products that have been reduced using the LAMOST\,2D reduction pipeline \citep[see,][]{2015RAA....15.1095Luo}, including sky background subtraction, wavelength calibration, flux calibration, and correction for telluric absorption in selected wavelength regions \citep{2015RAA....15.1095Luo,2017RAA....17...91Bai}. The spectra of Star\,1 and Star\,2 are of high quality, with SNR > 100 in the g, r and i bands, satisfying the minimum threshold of SNR > 10 adopted as a quality criterion for reliable spectroscopic analysis. The spectra cover a wavelength range of 3690\,--\,9000\,{\AA} at a spectra resolution of R\,$\sim$\,1800 \citep{2012RAA....12.1197Cui}. We continuum normalized the spectra using standard IRAF tasks \citep{1986SPIE..627..733Tody,1993ASPC...52..173Tody}.

\subsection{Optical: Timeseries photometry}\label{sec:timeseriesdata}
\subsubsection{TESS data reduction}\label{sec:TESSdatareduction}
We searched for TESS Target Pixel Files (TPFs) of the two program stars using the function \texttt{search\_targetpixelfile} \citep{2019AJ....157...98Ginsburg} available within the \textsc{python} package \textsc{lightkurve} \citep{2018ascl.soft12013Lightkurve}. TPFs were found only for Star\,1, while for Star\,2, only Full Frame Images (FFIs) were available. However, considering the large angular resolution of TESS ($\sim$21\,arcsec), care must be taken to check for flux contamination from neighboring sources as our program stars are found within the FOV of the cluster NGC\,1893 (see Figure \ref{fig:clusterNGC1893}). To check the amount of flux contamination for both the stars in our sample, we calculated the contamination ratio, which is the total flux from all the neighboring stars within $\sim$1\,arcmin divided by the flux of the target star. For this we used the python code \texttt{tic\_inspect}\footnote{\url{https://github.com/mpaegert/tic_inspect}} as described in \cite{2018AJ....156..102Stassun} and \cite{2022AJ....163..226Labadie-Bartz}. We find a contamination ratio of $\sim$\,10\% for Star\,1 and $\sim$\,46\% for Star\,2. Further, we visually checked the 1\,arcmin\,$\times$\,1\,arcmin FOV around both stars for any neighboring stars within or near the adopted aperture for light curve extraction that are less than 5 magnitudes fainter in the Gaia G-band filter. No such sources were found around Star\,1. In contrast, Star\,2 is located in a crowded field near SIM\,129 (see Figure \ref{fig:clusterNGC1893}), resulting in significant photometric contamination from nearby sources. Therefore, we extracted the TESS lightcurves only for Star\,1. 

Star\,1 in our sample was observed in seven sectors (43, 44, 45, 59, 71, 73, 86) from 2021 to 2024. The lightcurves were extracted from the 2-minute cadence TPFs. We used the aperture provided by the Science Processing Operations Center (SPOC) pipeline for light curve extraction. While the TPFs are corrected for background by the pipeline, some long-term trends other than scattered light, such as noise and spacecraft motion, could be present, which need to be corrected. We employed PCA-based de-trending to suppress instrumental systematics while preserving intrinsic stellar variability. Specifically, we used the \textsc{lightkurve} tool \texttt{RegressionCorrector}, which models the light curve systematics using a design matrix constructed from pixel-level data. Principal Component Analysis (PCA) is applied to reduce the matrix to its top five components, sufficient to capture trends caused by scattered light and spacecraft motion. This design matrix is then applied to the raw light curve to obtain a clean corrected light curve. The TESS light curve, originally in flux units (e$^{-}$\,s$^{-1}$), was converted to magnitudes using the relation $mag\,=\,-2.5\,log10(flux)\,+\,20.44$ from \cite{Vanderspek2018_tess_handbook}. 

\subsubsection{ZTF data preparation}\label{sec:ZTFdata}
We used the IRSA ZTF Lightcurve Service \footnote{\url{https://irsa.ipac.caltech.edu/Missions/ztf.html}} \citep{2019PASP..131a8002Bellm, 2019PASP..131a8003Masci} to search for the ZTF light curves of the two program stars. ZTF\,DR24 light curves were available only for Star\,2, observed from March 2018 to October 2025. Two sets of g-band and r-band light curves, and one i-band light curve were available. The two sets of g- and r-band light curves represent observations recorded on two different CCDs at the same time. However, the source was not consistently visible on one of the CCDs, giving rise to significant data gaps. Hence, we used only one of the two available sets of g- and r-band data. The i-band light curve could not be used as it had very few data points, making it unreliable for period analysis. Further, for period analysis of the g- and r-band light curves, we considered only the best quality data points with \texttt{catflags\,=\,0}. 

\section{Characterization of the program stars}\label{sec:analysisandresults}
\subsection{Spectral type estimation}\label{sec:sptestimation}
The spectral type was reported only for Star\,1 by \cite{2021RAA....21..288Shridharan}, whereas Star\,2 could not be classified due to complex spectral features. The LAMOST\,DR8 has good-quality data (SNR in g, i, and r-bands \textgreater 100) for both the stars, with updated flux calibration and pipeline reprocessing. Hence, we used the spectra from LAMOST\,DR8 and re-estimate the spectral type of the two program stars using similar methods mentioned in \cite{2021MNRAS.501.5927Anusha} and \cite{2021RAA....21..288Shridharan}. The spectral type of the two program stars was determined through a template-matching technique, utilizing the templates from the MILES spectral library \citep{2006MNRAS.371..703MILES} as its resolution is comparable with LAMOST. This library contains spectra from approximately 980 stars, spanning a wavelength range of 3525\,--\,7500 {{\AA}} and a spectral resolution of 2.5 {{\AA}} \citep{2011A&A...532A..95Falcon}, all acquired using the Isaac Newton Telescope (INT) situated at La Palma in the Canary Islands, Spain. The template stars encompass a broad parametric range, specifically T$_\mathrm{eff}$\,=\,3000\,--\,40000\,K and log (g)\,=\,0.2\,--\,5.5. First, we matched the templates from the spectral library with our sample using an automated Python code based on three absorption lines, i.e., H$\zeta$\,3889\,{\AA}, H$\varepsilon$\,3970\,{\AA}, and H$\delta$\,4101.7\,{\AA}. This allowed us to constrain the number of templates to match, providing only the best-fit templates. To achieve improved control over the spectral type, we visually examined the spectra to ensure that the lines used for classification were clear of any noise. In particular, the temperature-sensitive He\textsc{i} and Mg\textsc{ii} absorption lines were inspected along with the overall spectral features. The best-fit MILES template that closely matched the observed spectrum was assigned as the spectral type of the program stars. We derived a spectral type of B5V and B7IV for the program stars 1 and 2, respectively, with an uncertainty of two subclasses due to the limited spectral resolution \citep{2021RAA....21..288Shridharan}. Earlier classifications based on photometry and very low-resolution spectroscopy \citep{1966IAUS...24..373Hiltner, 2007A&A...471..485Negueruela} are also consistent with our results. We tabulate our findings in Table \ref{tab:qparamtab}. The spectral features of the two stars are discussed individually in section \ref{sec:star1CXOUJ052218.4+332820} and section \ref{sec:star2CXOUJ052304.2+332846}.

\begin{table*}[ht]
    \centering
    \caption{Details of the stellar parameters of the two target stars derived in this study, comprising spectral type from template fitting, distances in pc from \cite{BailerJones2018}, A$_{\rm V}$, R$_{\star}$, and L$_{\rm bol}$ in solar units, and T$_{\rm eff}$.}  
    \small
    \setlength{\tabcolsep}{3.25pt}
    \renewcommand{\arraystretch}{1.15}
    \begin{tabular}{cccccccccc}
    \hline 
        \textbf{Sl. No} & \textbf{Object} & \textbf{RA [J2000]} & \textbf{Dec [J2000]} &
        \textbf{Dist (pc)} & \textbf{SpT} & \textbf{A$_\mathrm{V}$} &
        \textbf{logL$_\mathrm{bol}$ (L$_{\odot}$)} &
        \textbf{T$_\mathrm{eff}$ (K)} &
        \textbf{R$_{\star}$ (R$_{\odot}$)} \\
        \hline \hline
        1 & CXOU\,J052218.4+332820 & 80.576859 & 33.472536 &
        1809.71$\pm$110.53 & B5V & 1.17 &
        2.71$^{+0.08}_{-0.12}$ & 15700 &
        3.06$^{+1.16}_{-0.78}$ \\
        
        2 & CXOU\,J052304.2+332846 & 80.767759 & 33.479554 &
        2779.12$\pm$318.99 & B7IV & 2.16 &
        2.31$^{+0.12}_{-0.25}$ & 14000 &
        2.43$^{+2.32}_{-0.98}$ \\
    \hline
    \end{tabular}
        \label{tab:qparamtab}
\end{table*}

\subsubsection{Estimation of A$_{V}$ from the spectral type}\label{sec:avfromtemp}
As discussed in the section \ref{sec:introduction}, NGC\,1893 exhibits significant differential extinction, with $\mathrm{E(B - V)}$\,$\sim$\,0.4\,--\,0.7\,mag \citep{2007MNRAS.380.1141Sharma, 2022MNRAS.515.3352Bijas}. Hence, we did not adopt the average extinction of the cluster, but rather estimated A$_\mathrm{V}$ for the program stars from the derived spectral type. The intrinsic color index $\mathrm{(B-V)_0}$ for the derived spectral type was obtained from \cite{PnM2013}, while the observed color index $\mathrm{(B - V)}$ was taken from \cite{1995ApJ...454..151Massey}. The color excess $\mathrm{E(B - V)}$ was then computed, and A$_\mathrm{V}$ was derived assuming a standard Galactic extinction law with R$_\mathrm{V} = 3.1$. The derived A$_\mathrm{V}$ values are 1.17 and 2.16 for Star\,1 and Star\,2, respectively (see Table \ref{tab:qparamtab}). From the estimated spectral type and A$_{\rm V}$, we calculated the effective temperature (T$_{\rm eff}$), stellar radii (R$_{\star}$), and bolometric luminosities (L$_{\rm bol}$). The T$_{\rm eff}$ and bolometric corrections corresponding to the derived spectral types were adopted from \cite{PnM2013}, while the R$_{\star}$ and L$_{\rm bol}$ were derived using the extinction-corrected magnitudes and Gaia distances. The derived stellar parameters are listed in Table \ref{tab:qparamtab}.

\subsection{Evolutionary status of the program stars}\label{sec:evolutionarystatusofprogramstars}
\subsubsection{Membership status}\label{sec:clustermember}
\begin{figure}[!ht]
    \centering
    \subfloat{\includegraphics[width=0.5\columnwidth]{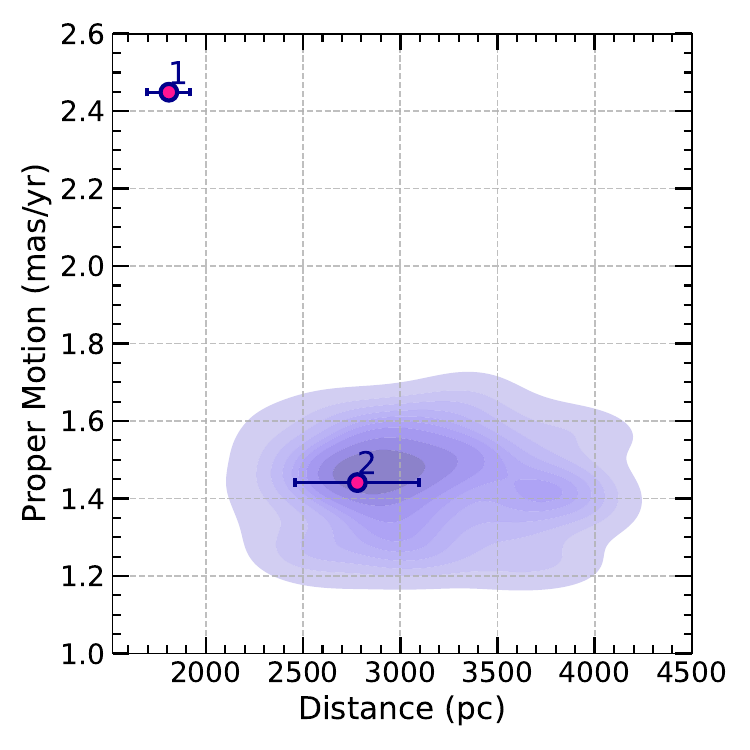}}
    \subfloat{\includegraphics[width=0.5\columnwidth]{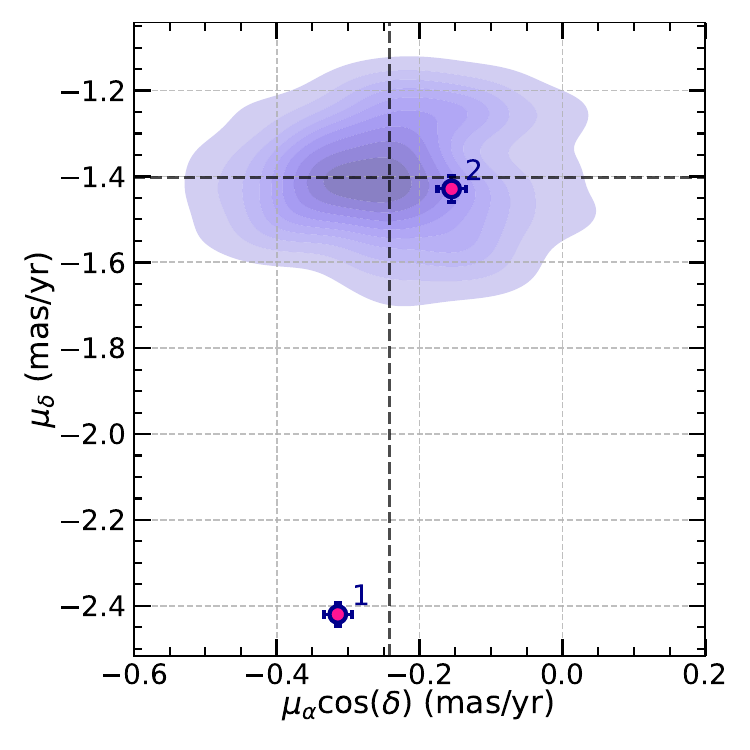}}
    \caption[A]{Gaia DR3 proper motion and distance membership analysis. (left) - distance against total proper motion. (right) - proper motion vector point diagram ($\mu_\alpha cos \delta$ vs. $\mu_{\delta}$). The program stars in our work are marked with solid pink circles. Confirmed members of the cluster NGC\,1893 from \cite{2024A&A...686A..42Hunt} with membership probability > 90\% are plotted as a blue contour.}
    \label{fig:clustemembp}
\end{figure}

The two program stars lie within the FOV of the cluster NGC\,1893 (Figure \ref{fig:clusterNGC1893}), raising a question as to whether they belong to the cluster or not. The membership probability for the two stars from the previous studies provided an uncertain status \citep{2019MNRAS.482..658Xue, 2018MNRAS.478.5184Dias}. Hence, we used the Gaia\,DR3 proper motion \citep{2023A&A...674A...1GaiaCollab} and distance \citep{BJ2021} to evaluate their cluster membership status. Based on our analysis, only Star\,2 lies within the 3$\sigma$ locus defined by confirmed members of the cluster NGC\,1893 \citep{2024A&A...686A..42Hunt}, while Star\,1 appears as an outlier. This suggests that Star\,2 is more likely to be physically associated with the cluster, whereas Star\,1 is a foreground field star. The proper motion plots from our study are presented in Figure \ref{fig:clustemembp}.

\subsubsection{2MASS color-color diagram}\label{sec:2massccd}
\begin{figure}[!ht]
    \centering
    \subfloat{\includegraphics[width=0.5\columnwidth]{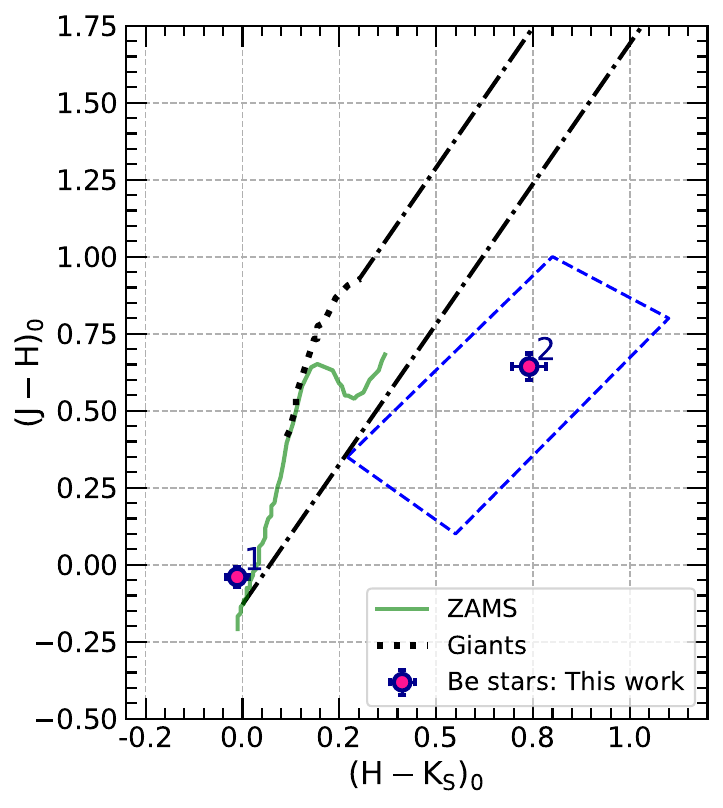}}
    \subfloat{\includegraphics[width=0.5\columnwidth]{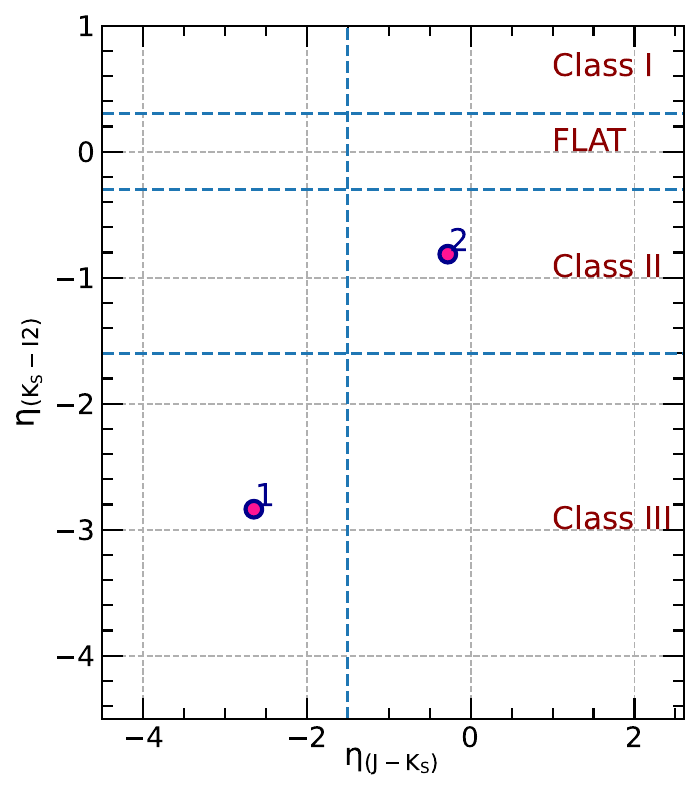}}
    \caption[A]{Left: 2MASS $\mathrm{(J-H)_0}$ versus $\mathrm{(H-K_{S})_0}$ color–color diagram of the two ELSs Star\,1 (CXOU\,J052218.4+332820) and Star\,2 (CXOU\,J052304.2+332846). The solid, dotted, and dot-dashed lines represent the main-sequence, the giant branch, and the reddening vectors, respectively, adopted from \cite{1983A&A...128...84Koornneef}, converted to the 2MASS system using the relations from \cite{2001AJ....121.2851Carpenter}. The blue box with dashed lines represents the region occupied by HAeBe stars from \cite{2005AJ....129..856Hernandez}. Right: The NIR and MIR spectral index plot. The blue dotted line represents excess in the NIR and MIR regions.}
    \label{fig:IRCCD+specindex+HR}
\end{figure}

The two ELSs in our sample are not well characterized in previous studies, with discrepancies in their object classification and spectral types. To understand better, we plotted the 2MASS color-color diagram (CCDm) for these stars. We queried for our sample of two ELSs in the 2MASS point source catalog \citep{2003yCat.2246....0Cutri}. The J, H, K$_\mathrm{S}$ magnitudes obtained were corrected for interstellar extinction, where the extinction in each band was derived from A$_\mathrm{V}$ (see Table \ref{tab:qparamtab}) using the extinction law from \cite{1990ARA&A..28...37Mathis}. The reddening corrected 2MASS CCDm of the two program stars is shown in Figure \ref{fig:IRCCD+specindex+HR}. While Star\,1 has no IR excess and lies near the MS region, Star\,2 shows significant IR excess and falls within the HAeBe loci defined by \cite{2005AJ....129..856Hernandez}. These findings suggest Star\,1 to be an early-type MS star and Star\,2 to be a PMS star.
\subsubsection{NIR/MIR spectral index}\label{sec:NIRspectralindex}
It is known that the excess infrared flux observed in PMS stars is attributed to dust present in their circumstellar disk \citep{Manoj2006, Chen2016}. \cite{lada1987} demonstrated that the infrared spectral index is a crucial factor for identifying the presence of disks in young stellar objects (YSOs). In this study, we employed the infrared spectral index $\eta_{(2 - 4.6)}$ as a measure of infrared excess. \cite{1994Greene} defined a classification system for YSOs based on the slope of the SEDs in the infrared region. YSOs are categorized into four classes: Class I ($\eta$\,$\geq$\,0.3), Flat (-0.3\,$\leq$\,$\eta$\,<\,0.3), Class II (-1.6\,$\leq$\,$\eta$\,<\,-0.3), and Class III ($\eta$\,$\leq$\,-1.6). We computed the spectral indices $\eta_\mathrm{(J\,-\,K_S)}$ and $\eta_\mathrm{(K_S\,-\,I2)}$ for the two program stars (See Figure \ref{fig:IRCCD+specindex+HR}) using the method described in \citet{2011Manoj} and \citet{Arun_2019}. The J and K$_\mathrm{S}$ magnitudes were obtained from \cite{2003yCat.2246....0Cutri} and the Spitzer IRAC Ch2 (I2) from \cite{2013ApJS..209...27Kuhn}. The observed magnitudes were reddening corrected following the same procedure described in section \ref{sec:2massccd}. Star\,1 shows no significant IR excess with $\eta_\mathrm{(K_S\,-\,I2)}$\,=\,-2.84\,$\pm$\,0.04, consistent with a disk-less main-sequence B-type star. In contrast, Star\,2 exhibits strong NIR and MIR excess with $\eta_\mathrm{(K_S\,-\,I2)}$\,=\,-0.82\,$\pm$\,0.05, indicating the presence of a dusty circumstellar disk.

\subsubsection{Spectral energy distribution}\label{sec:SEDtext}
To further investigate the IR characteristics of our program stars, we performed SED fitting using multi-wavelength photometry. We used the Virtual Observatory SED Analyzer tool \citep[VOSA4 version 7.5;][]{2008A&A...492..277Bayo} to fit the synthetic stellar spectra to the data. The photometric data for the SEDs of the two target stars were obtained from the Vizier portal service, using a search radius of 5 arcsec. Further, we filtered the photometric data to contain only UBV \citep{1995ApJ...454..151Massey}; R \citep{2004AAS...205.4815ZachariasNOMAD}; G, G$_\mathrm{BP}$, and G$_\mathrm{RP}$ \citep{2023A&A...674A...1GaiaCollab}; the 2MASS J, H, and K$_\mathrm{s}$ \citep{2003yCat.2246....0Cutri}; the Spitzer IRAC channels Ch1, Ch2, Ch3, and Ch4 \citep{2013ApJS..209...27Kuhn}; the WISE W1, W2, W3, and W4 \citep{2014yCat.2328....0CutriWISE}; Gaia synthetic photometry in Johnson-Kron-Cousins I$_\mathrm{c}$ \citep{2023A&A...674A...1GaiaCollab}. We used the VOSA SED analyzer to fit the model BT-NextGen \citep{2009ARA&A..47..481Asplund} to the SED by allowing the A$_\mathrm{V}$, log(g), and T$_\mathrm{eff}$ parameters to vary. The initial values of A$_\mathrm{V}$ were set to values presented in the Table \ref{tab:qparamtab}. The parameters obtained from the best-fit models are T$_\mathrm{eff}\,$\,=\,15000\,K and log(g)\,=\,3.5, for both the program stars, while the A$_\mathrm{V}$ values obtained are 1.4 and 2.2 for Star\,1 and Star\,2, respectively. These A$_\mathrm{V}$ values are consistent with those derived from the spectral type of the two program stars (Table\,\ref{tab:qparamtab}). The SED of Star\,1 does not exhibit any infrared excess, consistent with a MS classification. Star\,2 shows significant IR excess, a characteristic commonly observed in PMS stars. These results are in agreement with the 2MASS CCD and the NIR spectral index.
\begin{figure}[!ht]
    \centering
    \subfloat{\includegraphics[width=\columnwidth]{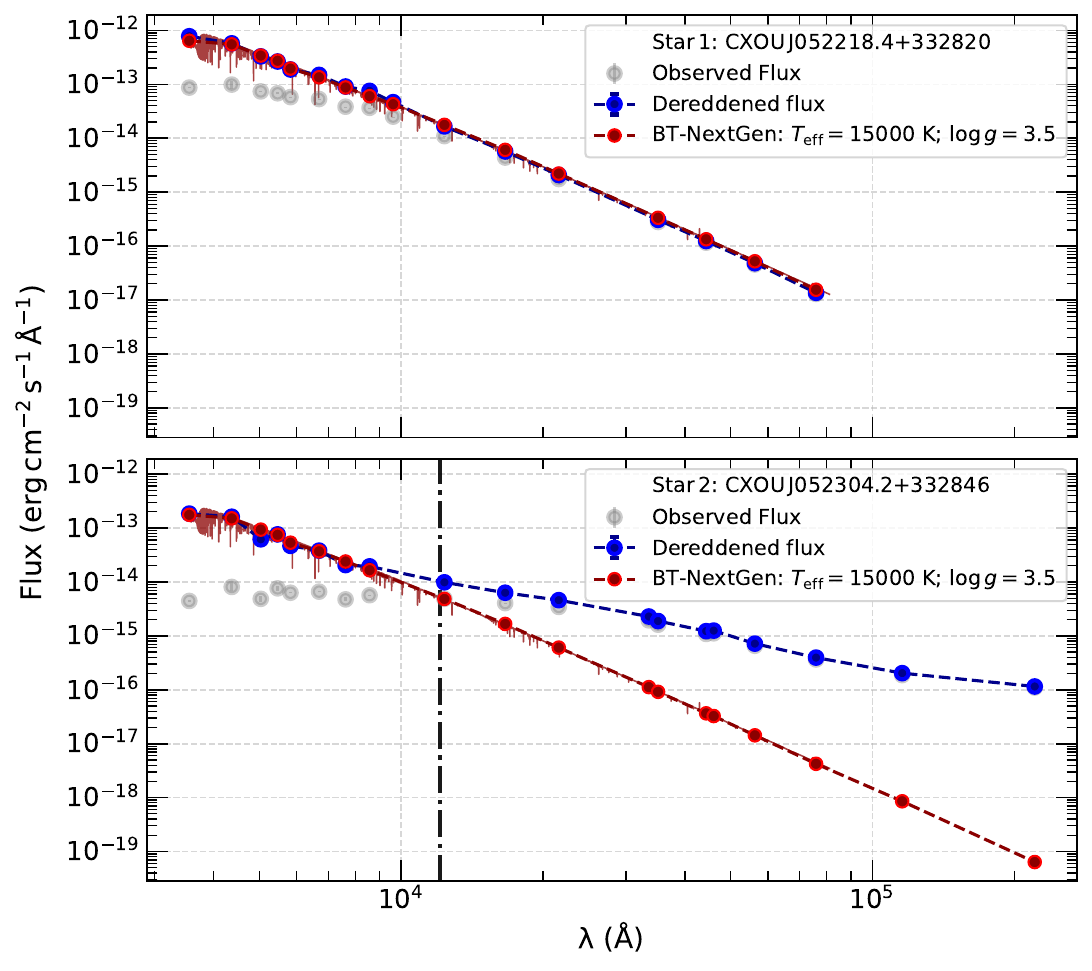}}\hspace{0.1mm}
    \caption[A]{SED of Star\,1 (CXOU\,J052218.4+332820) (top) and Star\,2 (CXOU\,J052304.2+332846) (bottom). The SEDs are fit with the BT-NEXTGen-AGSS2009 theoretical atmospheric model, overlaid in red.}
    \label{fig:SED1n2}
\end{figure}

\subsection{CXOU\,J052218.4+332820: A Classical Be star}\label{sec:star1CXOUJ052218.4+332820}
\subsubsection{Spectral features: LAMOST\,DR8\,LRS}\label{sec:star1spectralfeatures}

The spectral type of CXOU\,J052218.4+332820 (Star\,1) was estimated to be B5Ve based on our analysis of the LAMOST\,DR8 LRS (Section \ref{sec:sptestimation}). 

\begin{figure*}[!ht]
    \centering
    \subfloat{\includegraphics[width=0.9\linewidth]{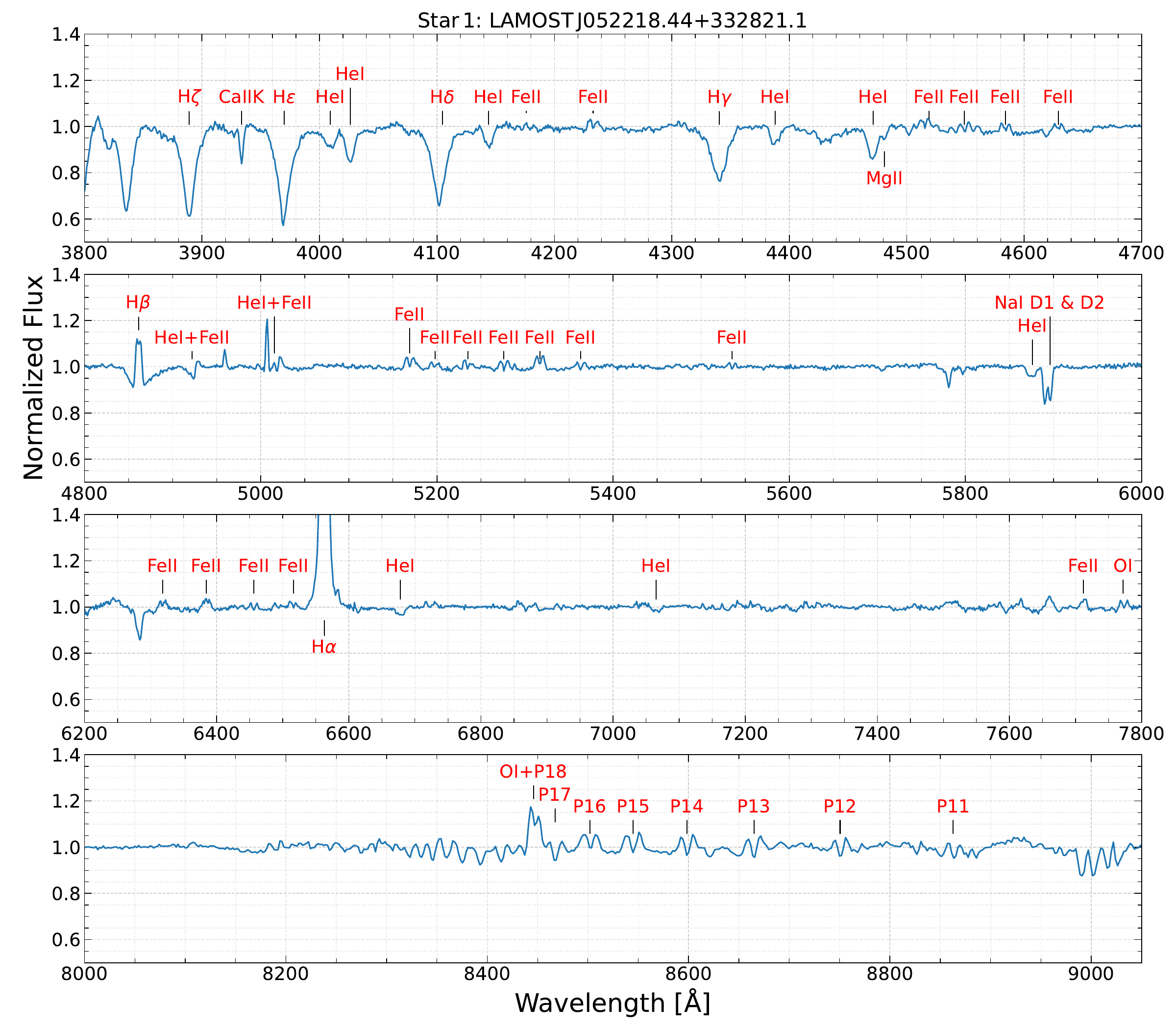}}
    \caption[A]{LAMOST\,DR8\,LRS spectrum of Star\,1 (CXOU\,J052218.4+332820). The prominent spectral lines are identified in red.}
    \label{fig:LAMOSTspec1}
\end{figure*}

\begin{figure}[!ht]
    \centering
    \subfloat{\includegraphics[width=0.7\columnwidth]{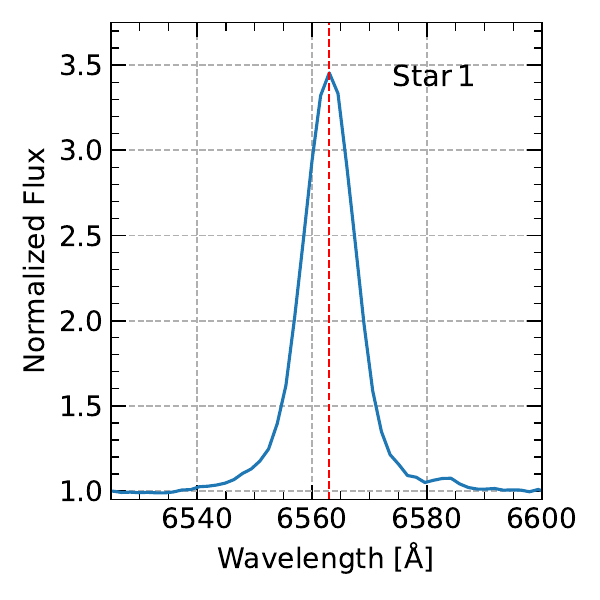}}
    \caption[A]{The H$\alpha$ line profile of Star\,1 (CXOU\,J052218.4+332820), depicting a symmetrical profile, with an EW of -35$\pm$1 {{\AA}} corrected for photospheric absorption. The red dashed line indicates the rest wavelength of H$\alpha$.}
    \label{fig:Object1halphaprofile}
\end{figure}

The spectrum exhibits several notable features such as the emission lines of H$\alpha$, H$\beta$, and various multiplets of Fe\textsc{ii} lines, O\textsc{i}, and the Paschen series (7500\,--\,9000\,{\AA}). The equivalent width (EW) of Balmer emission lines of hydrogen, i.e., H$\alpha$ and H$\beta$, is -35.0\,$\pm$\,1.0\,{\AA}, and -8.2\,$\pm$\,0.4\,{\AA}, respectively, consistent with the values observed for CBe stars \citep{2011BASI...39..517Mathew,2021MNRAS.500.3926Banerjee}. We measured the EW of the emission lines using standard \textsc{iraf} and corrected for underlying photospheric absorption using the best-fit MILES template for the program stars. The uncertainty in the EW measurements is $\sim$\,10\%. The H$\alpha$ profile is single-peaked and symmetric (see Figure \ref{fig:Object1halphaprofile}), whereas the H$\beta$ line displays a weak double-peaked structure (i.e., absorption-in-emission). The single-peaked appearance of the H$\alpha$ line may be a result of the low-resolution spectrum \citep{2021MNRAS.500.3926Banerjee}. Apart from the aforementioned features, the spectrum also displays double-peaked emission lines of Fe\textsc{ii}, O\textsc{i}, and Paschen series, indicating their origin in a rotating circumstellar disk. \cite{1993A&AS...97..807Jaschek} using the Keplerian disk model proposed by \cite{1972ApJ...171..549Huang}, estimated the radial distance of the line-emitting regions in Be stars in units of stellar radii (R$_{\star}$), based on the observed velocity separation between the violet and red components of emission line profiles. Following the same approach, we estimated the radial distances of emission regions of lines showing double peaks in the LAMOST\,DR8 spectrum using the equation from \cite{1972ApJ...171..549Huang}\citep[e.g.,][]{2016RAA....16...76Bhat,2012ApJ...753...13Mathew,2017JApA...38....6Paul}. The equation is given below, 
\begin{equation}
    \frac{R_d}{R_{\star}} = \left(\frac{2 \, \nu\,sin\,i}{\Delta V}\right)^{1/j}
    \label{eqn:diskregion}
\end{equation}
where R$_\mathrm{d}$ is the disk radius estimated in terms of R$_{\star}$, $\Delta$V is the peak separation in velocity space between the blue and red components of a double-peaked emission line. For a Keplerian orbit, j\,=\,1/2. The rotational velocity of the star ($\nu\,\mathrm{sin\,i}$\,=\,404$\pm$14\,km\,s$^{-1}$) was obtained from \cite{2010ApJ...722..605Huang}. We estimated the emitting regions for H$\beta$, Fe\textsc{ii} multiplets, O\textsc{i}\,7772 {\AA}, and Paschen series (Pa11 to Pa16) lines to lie at radial distances of approximately $\sim$\,15\,R$_{\star}$, 3\,--\,5\,R$_{\star}$, $\sim$4\,R$_{\star}$, and 3\,--\,5\,R$_{\star}$, respectively. These values are consistent, within uncertainties, with the mean radii of emission regions from \cite{1988A&A...189..147Hanuschik}, \cite{2006A&A...460..821Arias}, \cite{1993A&AS...97..781Jaschek} and \cite{1990A&AS...84...11Andrillat} for H$\beta$, Fe\textsc{ii} multiplets, O\textsc{i}\,7772\,{\AA} and Paschen series, respectively. Similarly, \cite{1992ApJS...81..335Slettebak} reported an estimate of $\sim$\,7\,--\,19\,R$_{\star}$ for the H$\alpha$ line. We could not derive this parameter for H$\alpha$ as it only exhibits a single-peak profile. It is evident that the Fe\textsc{ii}, O\textsc{i} 7772\,{\AA}, and Paschen series emission lines originate in the inner regions of the circumstellar disk, while the hydrogen lines, i.e., H$\alpha$ and H$\beta$, form farther out in the disk \citep{2012ApJ...753...13Mathew}. We do not observe the Ca\textsc{ii} triplets in the spectrum of this object, consistent with these lines being only occasionally seen in CBe stars \citep{2011BASI...39..517Mathew,2021MNRAS.500.3926Banerjee}. Star\,1 was also included in the sample of rapidly rotating B-type stars analyzed by \citet{2010ApJ...722..605Huang}. They suggested that some of the rapid rotators \citep[see, Table 8 in][]{2010ApJ...722..605Huang}, lacking emission at the time of their study, may develop circumstellar disks in the future. Photometric studies by \cite{1991MNRAS.253..649Tapia} and \cite{2001AJ....121.2075Marco} showed Star\,1 to have a $\beta$\,index\,>\,2.55, indicating that H$\beta$ line was not in emission during the observational epoch. The presence of prominent emission features such as H$\alpha$, H$\beta$, Fe\textsc{ii}, and Paschen series in the LAMOST\,DR8 spectrum confirms that Star\,1 currently hosts a gaseous disk. This suggests that Star\,1 has transitioned into an active, transient CBe phase, consistent with the interpretation made by \cite{2010ApJ...722..605Huang} for near-critical rotators. The observed double-peaked emission profiles, characteristic of a rotating circumstellar disk viewed at moderate-to-high inclination angles \citep[$i$\,>\,40\textdegree;][]{2017A&A...601A..74Klement}, together with the inferred disk radii and the transient nature of the disk, are consistent with the properties of CBe stars.

\subsubsection{Variability analysis: TESS lightcurves}\label{sec:variabilityanalysistess}

\begin{figure}[!ht]
    \centering
    \subfloat{\includegraphics[width=\columnwidth]{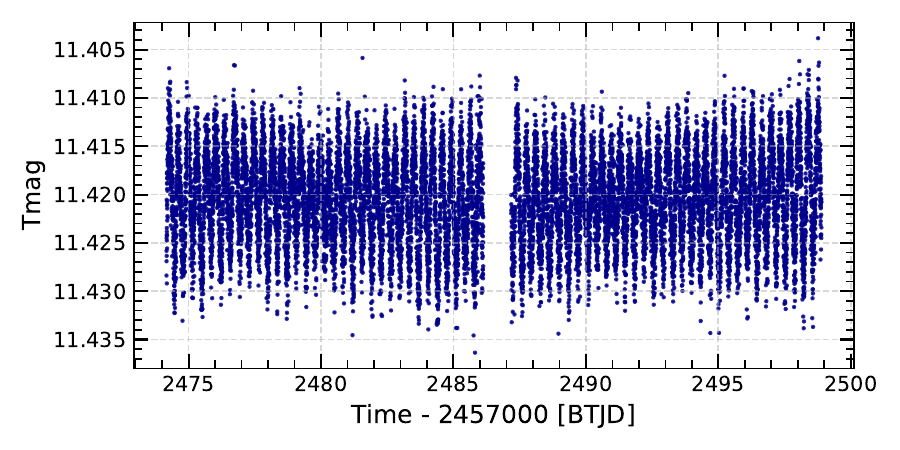}}\hspace{0.01mm}
    \subfloat{\includegraphics[width=\columnwidth]{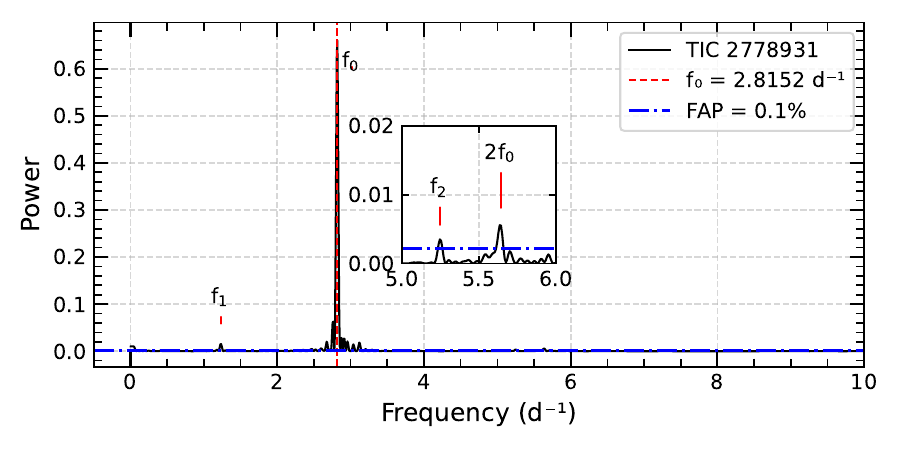}}\hspace{0.01mm}
    \subfloat{\includegraphics[width=\columnwidth]{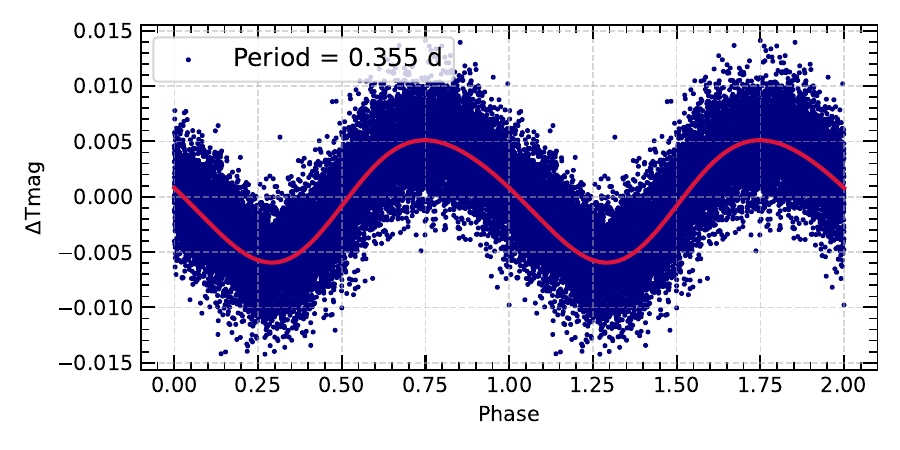}}
    \caption[A]{(Top) - TESS lightcurve for Star\,1 (CXOU\,J052218.4+332820) from section 43. (Middle) - The Lomb-Scargle periodogram derived for Star\,1. The dominant frequency and its 1st harmonic are denoted as f$_0$ and 2f$_0$, respectively. f$_1$ and f$_2$ are independent frequencies. (Bottom) - The phase-folded light curve of Star\,1, folded at the period of $\sim$\,0.35 days, derived from the dominant frequency f$_0$.} 
    \label{fig:TESS_TS_PGM_PFLC}
\end{figure}

We investigated the variability of the TESS light curves of Star\,1 across all seven sectors using the Lomb-Scargle periodogram \citep[LSP;][]{1976Ap&SS..39..447Lomb, 2009A&A...496..577Zechmeister}. For this, we used the class \texttt{LombScargle} from the \texttt{astropy.stats} \textsc{python} package. Prior to this, the TESS light curves were detrended and normalized. The LSP is shown in the middle panel of the Figure \ref{fig:TESS_TS_PGM_PFLC}. From the periodogram, we identified a dominant frequency at f$_0$\,$\sim$\,2.82\,d$^{-1}$ (P$_0$\,$\sim$\,0.355\,d), its harmonic at 2f$_0$, and two independent frequencies at f$_1$\,$\sim$\,1.24\,d$^{-1}$ and f$_2$\,$\sim$\,5.25\,d$^{-1}$. Upon pre-whitening, we obtained three more frequencies at f$_3$\,$\sim$\,2.46, f$_4$\,$\sim$\,2.90, and f$_5$\,$\sim$\,3.11\,d$^{-1}$. The dominant frequency has a false alarm probability (FAP) less than 10$^{-6}$, indicating a very strong signal. Other frequencies identified are fairly above the FAP level of 0.1\,\% (see Figure \ref{fig:TESS_TS_PGM_PFLC}). The periods obtained from our analysis are consistent with the range of periods reported in earlier studies, namely 0.264\,d \citep{2008JPhCS.118a2078Zhang}, 0.30\,d \citep{2012MNRAS.427.1449Lata}, and P$_1$=0.355\,d and P$_2$=0.18\,d \citep{2019MNRAS.482..658Xue}. Table \ref{tab:TESS_frequencies} provides details of the dominant frequency and its harmonic, along with their corresponding amplitudes for all seven sectors. Both f$_0$\,$\sim$\,2.815\,--\,2.820\,d$^{-1}$ and 2f$_0$\,$\sim$\,5.560\,--\,5.636\,d$^{-1}$ exhibit variability from sector-to-sector. The amplitude of f$_0$ (A$_1$) varies from 4.96\,--\,7.61\,mmag and that of 2f$_0$ (A$_2$) varies from 0.50\,--\,0.92\,mmag. Such variability in amplitudes and frequencies over time is attributed to the rotational modulation observed in CBe stars occurring due to the evolution of co-rotating circumstellar structures. The phase-folded light curve at P$_0$ (see lower panel of Figure \ref{fig:TESS_TS_PGM_PFLC}) shows a single sinusoid-like curve with asymmetry i.e., the steepness of the rising curve is more than that of the falling curve. Such an asymmetry, along with amplitude variations is attributed to the rotational modulation occurring due to the co-rotating clouds or spots in CBe stars \citep{2018PASA...35...34Ozuyar,2020MNRAS.493.2528Balona}. 

To further support the rotational origin, we estimated the maximum possible rotational period (P$_\mathrm{rot}$) from the projected rotational velocity ($\nu\,\mathrm{sin\,i}$) using the equation \ref{eqn:rotperiod}. 
\begin{equation}
    P_{rot} \leq \frac{2\pi R_{\star}}{\nu\,sin\,i}
    \label{eqn:rotperiod}
\end{equation}
where, R$_{\star}$ is the stellar radius in km derived using L$_\mathrm{bol}$ and T$_\mathrm{eff}$ mentioned in Table \ref{tab:qparamtab}, and $\nu\,\mathrm{sin\,i}$ = 404$\pm$14\,km\,s$^{-1}$ was obtained from \cite{2010ApJ...722..605Huang}. From equation \ref{eqn:rotperiod} we obtained an upper limit for P$_\mathrm{rot}$\,$\leq$\,0.38\,$\pm$\,0.08\,d. This P$_\mathrm{rot}$ is consistent with Star\,1's dominant period P$_0$\,=\,0.355\,d identified from the TESS light curves. Considering the TESS period of 0.355\,d to be true rotational period, we obtain the equatorial velocity $\nu_\mathrm{eq}$\,$\sim$\,436\,$\pm$\,85\,km\,s$^{-1}$. This implies a sin\,i\,$\sim$\,0.93 and i\,$\sim$\,68\textdegree. From the uncertainties in R$_{\star}$ and $\nu$\,sin\,i, we obtain a range for i\,$\sim$\,50\,--\,75\textdegree. This inclination agrees with the observed LAMOST\,DR8 line profiles of Fe\textsc{ii}, O\textsc{i}, and Paschen series showing double peak, H$\beta$ with a central shallow dip, while H$\alpha$ remains a nearly symmetric single peak (Figure \ref{fig:LAMOSTspec1} and \ref{fig:Object1halphaprofile}). This is typical of Be disks viewed at moderately high inclinations. This confirms that the observed TESS variability is plausibly caused by rotational modulation. 

\subsection{CXOU\,J052304.2+332846: A Herbig Be star}\label{sec:star2CXOUJ052304.2+332846}
\subsubsection{Spectral features: LAMOST\,DR8\,LRS}\label{sec:spectralfeaturesstar2}

\begin{figure*}
    \centering
    \subfloat{\includegraphics[width=0.9\linewidth]{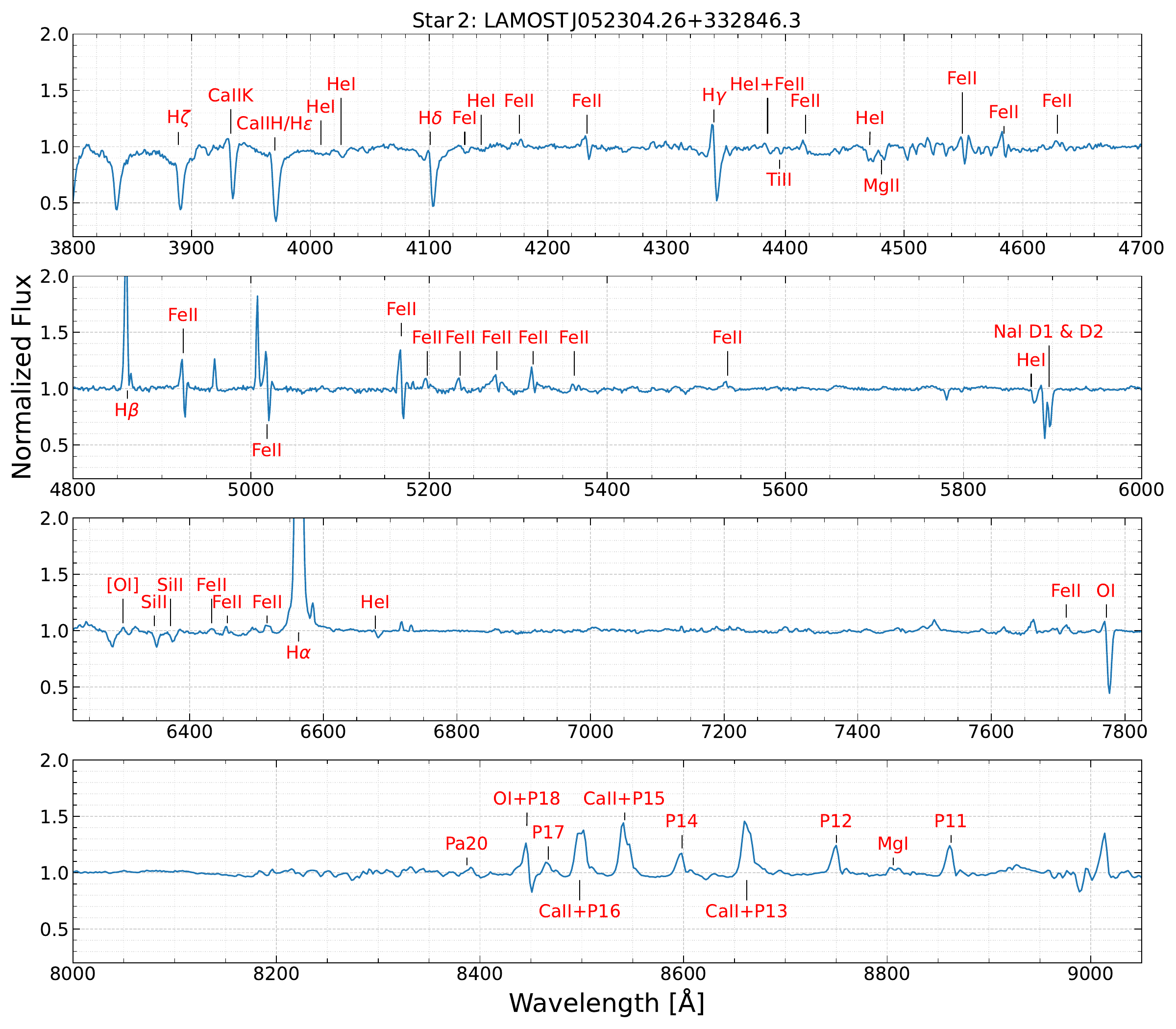}}
    \caption{LAMOST\,DR8\,LRS spectrum of Star\,2 (CXOU\,J052304.2+332846). The prominent spectral lines are identified in red.}
    \label{fig:LAMOSTspec2}
\end{figure*}

The spectral type of CXOU\,J052304.2+332846 (Star\,2) from the template fitting (Section\,\ref{sec:sptestimation}) was estimated to be B7IVe. Figure\,\ref{fig:LAMOSTspec2} displays only the parts of the LAMOST\,DR8 spectrum of Star\,2 with prominent spectral features. Upon visually examining the spectrum, we observe that the spectrum does not have any He$\textsc{ii}$ lines, and has very weak He$\textsc{i}$ lines present in the blue region ($\sim$\,4000\,{\AA}). The Balmer line profiles, particularly H$\zeta$, Ca\textsc{ii}\,H/H$\varepsilon$, and H$\delta$, exhibit accretion signatures. Therefore, the spectral subtype is primarily constrained based on the He$\textsc{i}$\,4471\,{\AA} and Mg$\textsc{ii}$\,4481\,{\AA} lines, which are well defined. These lines present approximately equal strength, consistent with a B7-type star \citep{2010A&A...517A..67Carmona}, although the accretion-related changes in the higher-order Balmer lines may introduce some uncertainty into the spectral classification. The luminosity class IV assigned in this work is consistent with the observed properties of HAeBe stars \citep{1998ARA&A..36..233Waters}. Furthermore, the surface gravity (log\,g\,$\sim$\,3.5) derived in our work from the SED fitting is consistent with the lower surface gravities expected for HAeBe stars undergoing PMS contraction toward the main sequence \citep{1972ApJ...173..353Strom,2013MNRAS.429.1001Alecian}. Finally, Star\,2 is a member of the very young cluster NGC\,1893 ($\sim$\,5\,Myr), confirming its PMS status. Thus, the derived stellar parameters and the evolutionary status of Star\,2 are consistent with its adopted HBe classification. However, it must be noted that the extinction derived in our work may include a contribution from the circumstellar disk, as circumstellar extinction can be significant in PMS stars \citep{2006A&A...456.1045Blondel}. 

\begin{figure}[!ht]
    \centering
    \subfloat{\includegraphics[width=0.7\columnwidth]{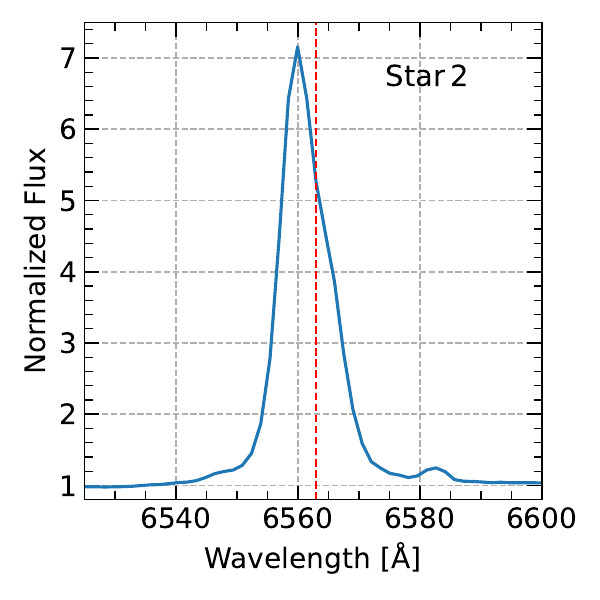}}
    \caption[A]{H$\alpha$ line profile of Star\,2 (CXOU\,J052304.2+332846). The red dashed line depicts the rest wavelength of H$\alpha$. The blue shift is evident for Star\,2 with an asymmetrical profile.}
    \label{fig:Object2halphaprofile}
\end{figure}

The LAMOST spectrum of Star\,2 (Figure \ref{fig:LAMOSTspec2}) exhibits clear evidence of accretion, i.e., gas infalling towards the star. This is observed in the complex line profiles of H$\alpha$, H$\beta$, H$\gamma$, H$\delta$, Ca\textsc{ii} K, H, and Fe\textsc{ii} lines. The spectral lines of H$\gamma$, and Fe\textsc{ii}\,(42) multiplets in our LAMOST spectrum have inverse P-Cygni (IPC) profiles \citep[Type \textsc{iv}-R;][]{1996A&AS..120..229Reipurth}, suggesting matter infall onto the star. The infall velocities of H$\gamma$ and the Fe\textsc{ii}\,(42) multiplets were estimated to be $\sim$\,130\,--\,150\,km\,s$^{-1}$. We observe the He\textsc{i} lines at 5876\,{\AA} and 6678\,{\AA} redshifted by $\sim$\,150\,--\,180\,km\,s$^{-1}$. These values are comparable to those reported for UX\,Ori-type stars \citep{1999A&A...343..137DeWinter}. The Si\textsc{ii} doublet $\lambda\lambda$ 6347/6371 are also redshifted. Other spectral lines, such as Ca\textsc{ii}\,K, Ca\textsc{ii}\,H/H$\varepsilon$, H$\delta$, and O\textsc{i}\,7772\,{\AA}, show IPC-like profiles with very weak emission. The detection of redshifted absorption in both Balmer (e.g., H$\gamma$) and metal lines (e.g., Fe\textsc{ii}) indicates that the infalling gas contains a significant hydrogen component and is consistent with a typical accretion episode occurring in HAeBe stars \citep{1994A&A...292..165Grinin,2000ApJ...542..421Natta}. The H$\beta$ shows a double-peak profile with V/R\,>\,1, consistent with a Type \textsc{iii}-R profile from the classification scheme introduced by \cite{1996A&AS..120..229Reipurth}. The H$\alpha$ line, on the other hand, shows strong emission (EW$_\mathrm{H\alpha}$\,=\,-67.3\,$\pm$\,2.9\,{\AA}) with an asymmetric single-peak profile, exhibiting a noticeable bump on the red (See Figure \ref{fig:Object2halphaprofile}). This case is consistent with the Type \textsc{ii}-R profile from \cite{1996A&AS..120..229Reipurth}. The spectrum also contains the forbidden line of [O\textsc{i}]\,6300\,{\AA} in faint emission. In the near-IR region of the spectrum of Star\,2 we observe spectral lines of O\textsc{i}, Ca\textsc{ii} triplets, and Paschen series, from Pa11 to Pa20, in emission with complex profiles. The Ca\textsc{ii} triplet lines at 8498, 8598, and 8662\,{\AA} are blended with the Paschen lines P16, P15, and P13, respectively. Other Paschen lines (P11, P12, P14, P17, and P20) posses Type \textsc{iii}-R kind emission lines similar to H$\beta$. The observed differences in the line profiles reflect their formation in different regions of the circumstellar environment. The He\textsc{i} lines, which show redshifted absorption, trace hotter and more compact regions associated with infalling gas close to the star. In contrast, the Paschen series, Ca\textsc{ii}, and lower Balmer lines (H$\beta$, and H$\alpha$) show stronger double-peaked emission, indicating a more extended line-forming region within the rotating circumstellar disk. The observed V/R\,>\,1 asymmetry in the double-peaked lines suggests that the circumstellar disk is not axisymmetric, i.e., it exhibits a non-uniform density distribution within the disk \citep{1994A&A...288..558Telting,1999A&A...343..137DeWinter}. From the H$\alpha$ emission line, we derive a mass accretion rate of ${\rm \dot{M}}_{\rm acc}$\,=\,1.99$^{+1.88}_{-0.97}$\,$\times$\,10$^{-6}$\,M$_{\odot}$\,yr$^{-1}$ using the empirical relations presented by \cite{Fairlamb2017}, consistent with values typically observed in actively accreting HAeBe and UX\,Ori-type systems \citep{2006A&A...459..837Garcia,2011Mendigutiab}. Similar line formation has been discussed in magnetospheric accretion models of other HAeBe stars, where compact accretion-related absorption features are partially or completely masked by emission originating from the extended inner disk \citep{2021Univ....7..489Pogodin}.

\begin{figure*}
    \centering
    \subfloat{\includegraphics[width=\textwidth]{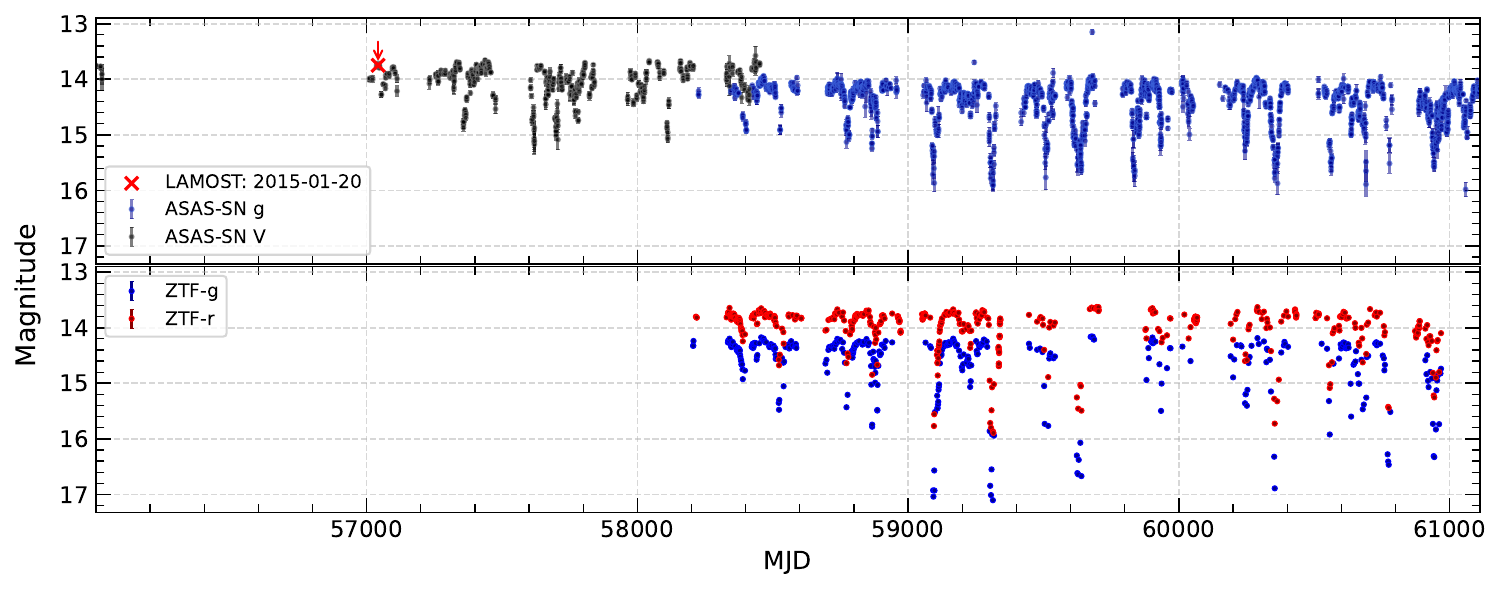}}
    \caption[A]{(Top) - ASAS-SN V-band and g-band light curves of Star\,2 (CXOU\,J052304.2+332846), (Bottom) - The ZTF\,DR24 ZTF-g and ZTF-r mag lightcurves of Star\,2, showing irregular dimming events.}
    \label{fig:ZTFDR24_LC}
\end{figure*}

HAeBe stars are known to show both spectroscopic and photometric variability \citep{1992PASP..104..479Graham, 1995AJ....110.2369Eaton, 1996A&A...309..474Grinin,1999A&A...343..137DeWinter}. The LAMOST spectrum of our program Star\,2, shows several signatures of a dynamically evolving circumstellar environment, including asymmetric emission-line profiles and accretion-related features. \cite{2007A&A...471..485Negueruela} using low-resolution spectrum obtained with Andalucia Faint Object Spectrograph and Camera (ALFOSC) on the 2.6\,m Nordic Optical Telescope (NOT) at La Palma, Spain, measured the EWs of H$\alpha$ and H$\beta$ of Star\,2 to be EW$_\mathrm{H\alpha}$\,=\,-55\,$\pm$\,2\,{\AA} \& EW$_\mathrm{H\beta}$\,=\,-3.7\,$\pm$\,0.3\,{\AA}, respectively. A difference of $\sim$\,12\,{\AA} in EW$_\mathrm{H\alpha}$ between the LAMOST (this work) and ALFOSC \citep{2007A&A...471..485Negueruela} suggests that Star\,2 is variable over time. This variability, along with the spectral features observed in our LAMOST spectrum of Star\,2, is consistent with behavior commonly seen in HAeBe stars, particularly those showing UX\,Ori-type characteristics, which are typically associated with moderate to high-inclination systems \citep{2007ApJ...656..980Pontoppidan,2013A&A...551A..21Kreplin,2014A&A...564A.118Vural,2016A&A...590A..96Kreplin}.

Star\,2 was observed for over $\sim$\,12\,years from 2014-12-19 to 2026-03-13 as part of the ASAS-SN survey. We retrieved the V-band and g-band light curves from the ASAS-SN Sky Patrol \citep{2014ApJ...788...48Shappee, 2017PASP..129j4502Kochanek}. The light curves are plotted in the top panel of Figure \ref{fig:ZTFDR24_LC}. The ASAS-SN light curves in both V- and g-bands show significant variability with deep irregular fadings. We see that the LAMOST spectrum of our program Star\,2 was observed during the bright phase on 2015-01-20. This, along with the spectral features discussed in this section, is similar to circumstellar and accretion signatures reported during the bright phases of UX\,Ori-type stars RR\,Tau, and UX\,Ori itself \citep{2000ApJ...542..421Natta,2002ApJ...564..405Rodgers}. Although our program Star\,2 has a long baseline ASAS-SN light curve, it is contaminated by bright neighboring sources due to the instrument's large pixel size ($\sim$\,8\,arcsec\,pixel; $\sim$\,15\,arcsec\,FWHM). Hence, we will not be using the ASAS-SN data for period analysis. In Section \ref{sec:variabilityanalysisZTF}, we investigate the photometric variability using ZTF light curves. 

\begin{figure*}
    \centering
    \subfloat{\includegraphics[width=\columnwidth]{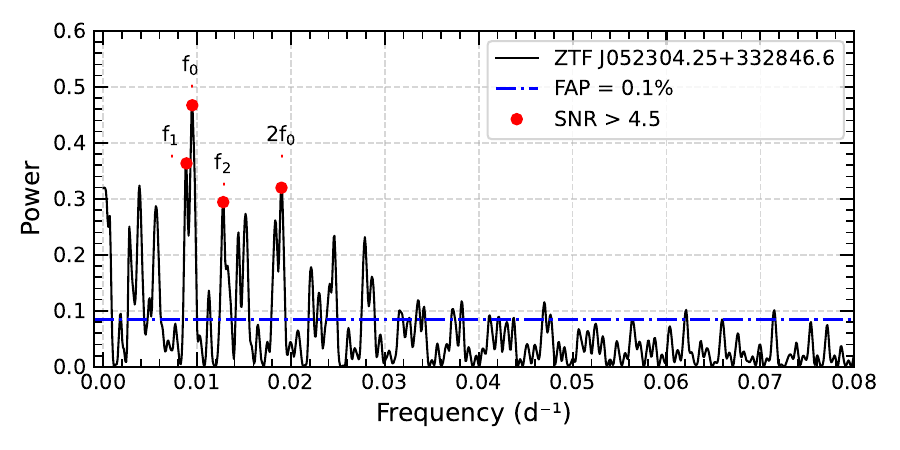}}\hspace{1mm}
    \subfloat{\includegraphics[width=\columnwidth]{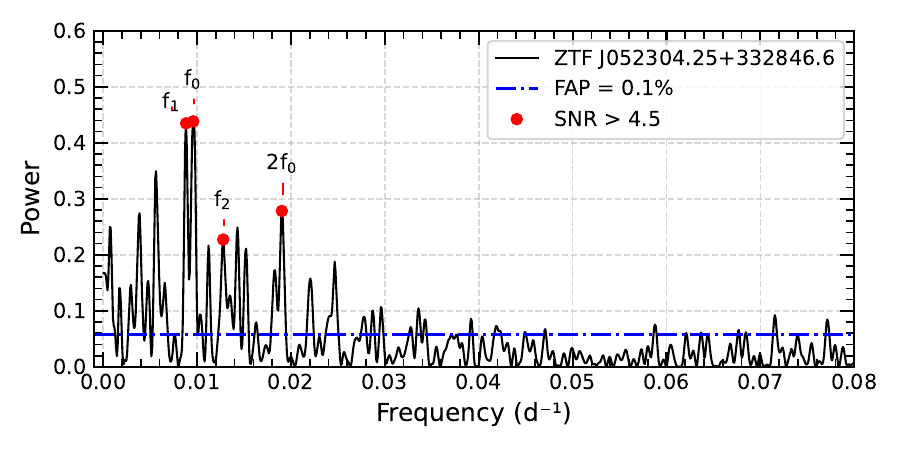}}\hspace{1mm}
    \subfloat{\includegraphics[width=\columnwidth]{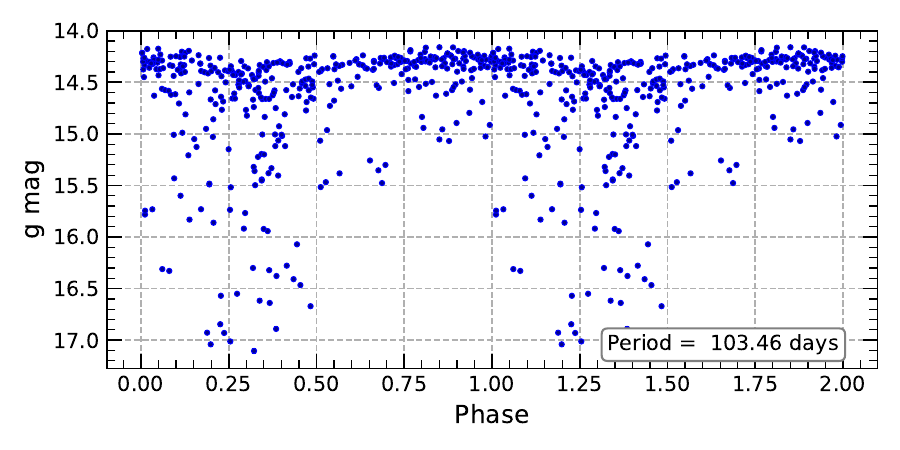}}\hspace{1mm}  
    \subfloat{\includegraphics[width=\columnwidth]{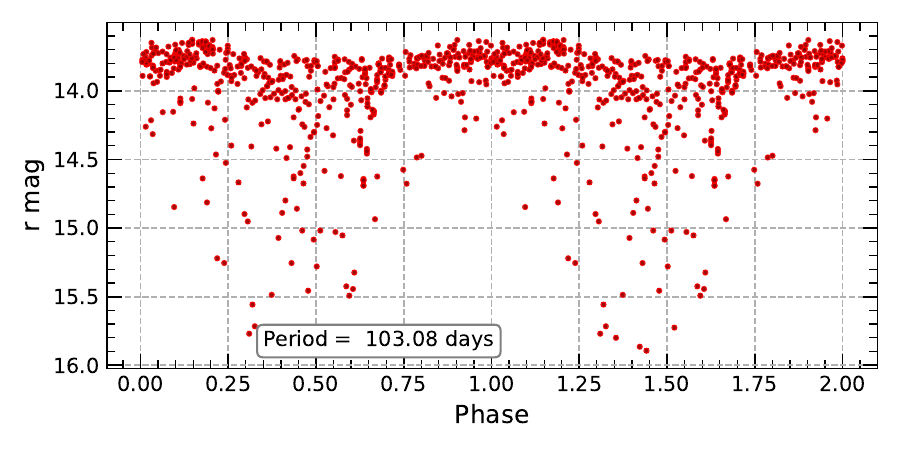}}\hspace{1mm} 
    \caption[A]{(Top) - Lomb-Scargle Periodogram of the ZTF-g mag light curve; (Bottom) - Phase-folded ZTF-g mag light curve at the best period (P$_0$) for Star\,2 (CXOU\,J052304.2+332846).}
    \label{fig:ZTFDR24_LSP+PF}
\end{figure*}

\subsubsection{Variability analysis: ZTF light curves}\label{sec:variabilityanalysisZTF}
The variability of ZTF light curves of Star\,2 was investigated in a similar manner to that of the TESS data mentioned in section \ref{sec:variabilityanalysistess}. The ZTF light curves of Star\,2, in g- \& r-band spanning from 2018-03-27 to 2024-10-30 ($\sim$\,7.5\,years) are shown in the bottom panel of Figure \ref{fig:ZTFDR24_LC}. The light curves exhibit irregular deep fading events ($\Delta$mag\,$\sim$\,3) driven by circumstellar extinction \citep{1998AstL...24..802Grinin}. We computed the LSP for ZTF-g \& ZTF-r light curves using the class \texttt{Lomb-Scargle} of \texttt{astropy.stats} \textsc{python} package. The LSP of Star\,2 (top panel of Figure \ref{fig:ZTFDR24_LSP+PF}) exhibits multiple broad and closely spaced peaks above the adopted FAP threshold (FAP\,=\,0.01\,\%). This is likely due to the quasi-periodic nature of the variability. Therefore, in addition to the FAP criterion, we consider only the frequencies with SNR\,>\,4.5 \citep{1993A&A...271..482Breger}. All of these frequencies identified have an FAP\,<\,10$^{-6}$ (See Figure \ref{fig:ZTFDR24_LSP+PF}), indicating strong signals. The four significant frequencies are listed in Table \ref{tab:ZTF_frequencies}. The dominant frequency at f$_0$\,=\,0.0096\,d$^{-1}$ and its first harmonic at 2f$_0$\,=\,0.0190\,d$^{-1}$, correspond to a period of $\sim$\,103\,d and $\sim$\,52\,d, respectively. Two other additional independent frequencies, f$_1$\,=\,0.0088\,d$^{-1}$ and f$_2$\,=\,0.0128\,d$^{-1}$ detected are associated with periods of $\sim$\,113\,d and $\sim$\,78\,d, respectively. The derived frequencies from the ZTF-g \& ZTF-r light curves (see Table \ref{tab:ZTF_frequencies}) are consistent within the uncertainties, suggesting that the observed variability is driven by coherent physical processes. The derived periods correspond to the intervals between successive dimming events seen in the light curve (bottom panel of Figure \ref{fig:ZTFDR24_LC}). We also note a weak long-timescale peak, corresponding to a variability period of $\sim$\,5\,years. However, its significance is too low to be considered statistically reliable, likely due to the limited temporal baseline. The bottom panel of Figure \ref{fig:ZTFDR24_LSP+PF} shows the light curves, phase-folded at the dominant frequency (f$_0$). The amplitudes corresponding to the dominant frequency are $\Delta$ZTF-g\,=\,2.945\,$\pm$\,0.023\,mag and $\Delta$ZTF-r\,=\,2.266\,$\pm$\,0.018\,mag, respectively. We see that the phase-folded light curve exhibits quasi-periodic variability associated with irregular circumstellar processes within dusty disk structures. The derived periods, corresponding to dimming timescales of several tens to hundreds of days, are consistent with the quasi-periodic variability commonly observed in UX\,Ori-type stars \citep{1991Ap&SS.186..283Grinin,1997ApJ...491..885Natta,1999AJ....118.1043Herbst}. We discuss this in detail in section \ref{sec:star2suxoritype}.

\begin{figure}
\centering
 \includegraphics[width=0.8\columnwidth]{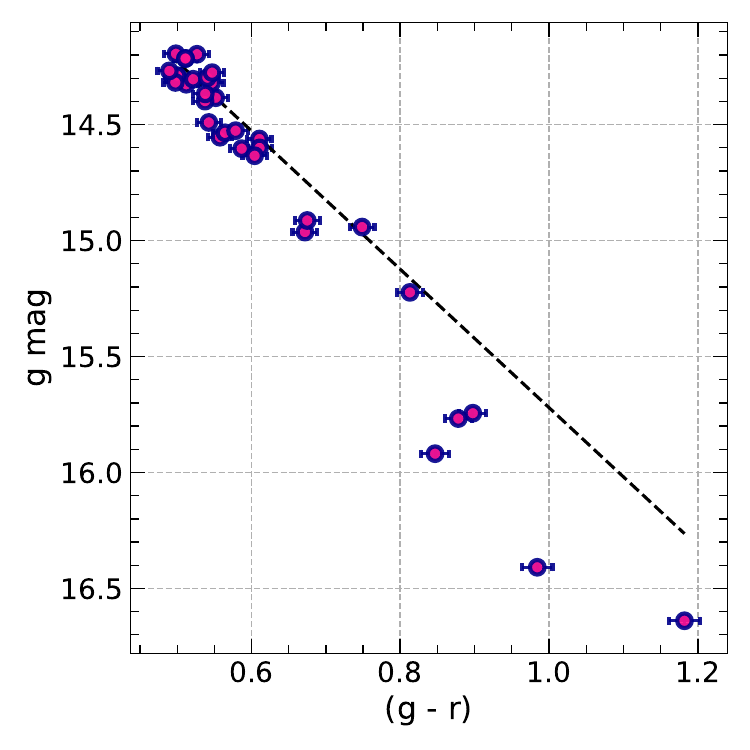}
   \caption{The CMD of Star\,2 (CXOU\,J052304.2+332846), a UX\,Ori-type HBe star, in ZTF-g versus (g-r), showing the color index becoming redder as the ZTF-g magnitude becomes fainter. The black dashed line represents the expected ISM reddening slope.} 
   \label{fig:blueingeffUXORI}
\end{figure} 

\subsubsection{UX-Ori\,type behavior in CXOU\,J052304.2+332846}\label{sec:star2suxoritype}

UX\,Ori-type stars are a subclass of HAeBe stars characterized by irregular deep photometric minima caused by variable circumstellar extinction from dusty structures within the inner circumstellar disk \citep{1991Ap&SS.186..283Grinin,1999AJ....118.1043Herbst}. One of the characteristic observational properties of these systems is the ``blueing effect'' \citep{1990A&A...236..155Bibo}, where the star becomes bluer during deep minima due to an increased contribution from scattered light when the star is obscured \citep{1988SvAL...14...27Grinin}. We investigate this behavior for Star\,2 by constructing color-magnitude diagram (CMD) using ZTF-g and ZTF-r photometry. The ZTF g- and r- light curve data were not observed simultaneously. Hence, we considered only measurements separated by less than $\sim$28\,minutes. The CMD is shown in Figure \ref{fig:blueingeffUXORI}. We do not observe a clear blueing effect in the CMD. One possible reason is that, due to the non-simultaneous nature of the photometric data, the deepest dimming events are not covered in the CMD. Further, we compared the observed reddening trend with the standard ISM extinction slope derived from the extinction law \citep{1990ARA&A..28...37Mathis}. The observed data points show a clear deviation from the expected ISM reddening slope, suggesting an additional circumstellar contribution to the extinction. The absence of the blueing effect in V-R versus V CMD has been observed in several UX\,Ori-type stars. Previous studies have shown that the strength of the blueing effect depends strongly on the relative contribution of scattered light, which is sensitive to the circumstellar geometry, disk inclination, dust distribution, and scattering angle \citep{1991Ap&SS.186..283Grinin}. For example, UX\,Ori exhibits a larger scattered-light contribution \citep[$\sim$10\%;][]{2000ASPC..219..216Grinin} than RR\,Tau \citep[$\sim$3\%;][]{1997A&A...327..145Rostopchina} in the optical bands. This difference was suggested as arising due to different inclination angle and scattering geometry of the circumstellar disk relative to the line of sight \citep{2025AA...694A.257Tambovtseva}.

As discussed in Section \ref{sec:variabilityanalysisZTF}, the ZTF light curves of Star\,2 exhibit irregular deep dimming events with amplitudes of $\Delta$ZTF-g\,$\sim$\,2.95\,mag and $\Delta$ZTF-r\,$\sim$\,2.26\,mag. The light curves also show multiple quasi-periodic frequencies (See Table \ref{tab:ZTF_frequencies}), corresponding to characteristic variability timescales of approximately 103, 52, 113, and 78\,d. This type of variability is consistent with the P$_\mathrm{1}$ quasi-periodic behavior discussed by \cite{1993Ap&SS.202..121Shevchenkopart1}. \cite{1993Ap&SS.202..137Shevchenkopart2} associated such variability with rotational modulation and non-stationary accretion-related structures producing variable circumstellar obscuration within the circumstellar disk. In addition, the observed spectral variability, strong circumstellar emission features, ongoing accretion activity, and deviations in the extinction properties from the standard ISM extinction curve further support the presence of a variable circumstellar environment. These observational characteristics suggest that Star\,2 is a strong UX\,Ori-type candidate. Further long-term simultaneous multi-band photometric and spectro-polarimetric observations would be useful to confirm the UX\,Ori nature of the source.

\begin{figure*}[!ht]
    \centering
    \includegraphics[width=0.8\linewidth]{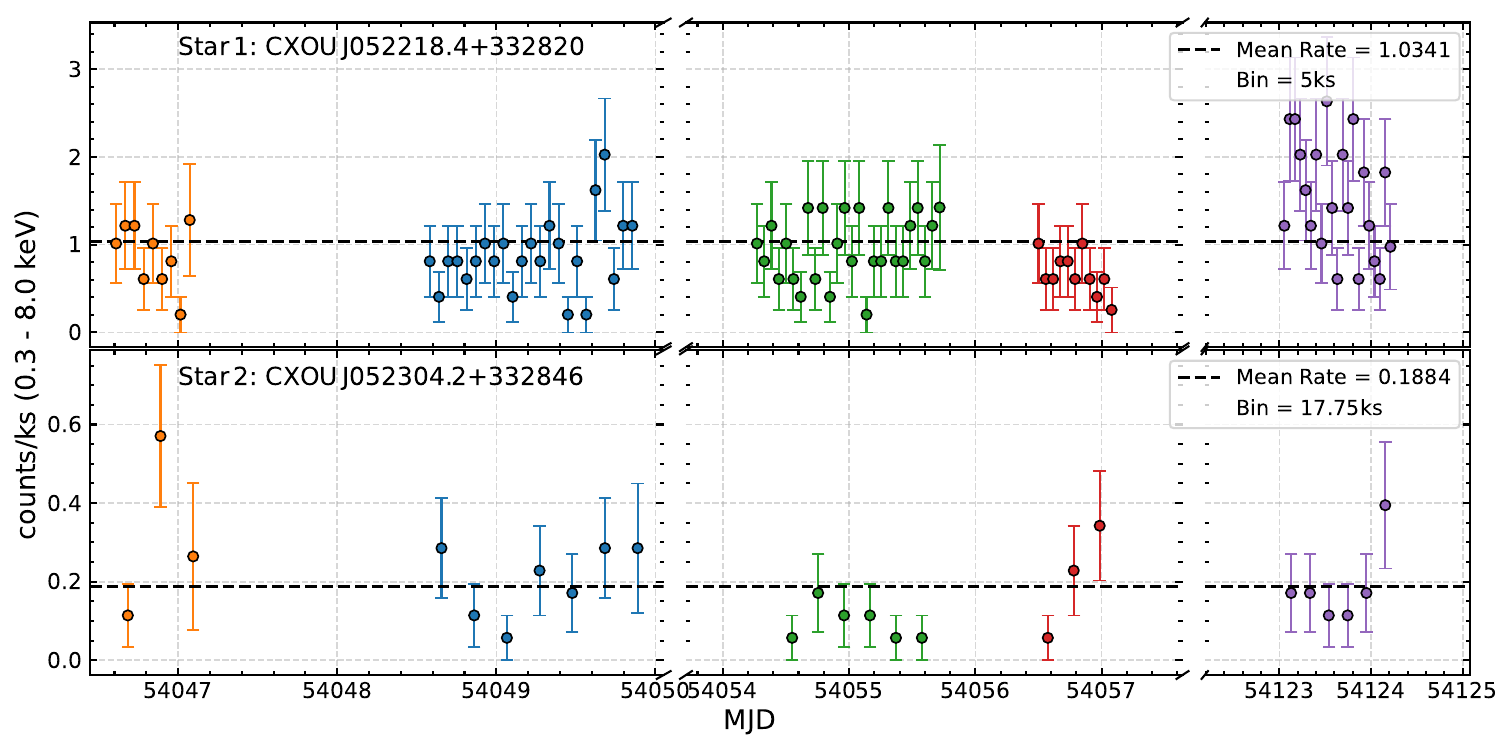}
    \caption{{\textit{Chandra}} X-ray lightcurves of (Top) Star\,1 (CXOU\,J052218.4+332820) and (Bottom) Star\,2 (CXOU\,J052304.2+332846). Both sources were observed at the same time from 2006--11--07 to 2007--01--24. No variability is observed for either of the sources, resembling a quiescent state.}
    \label{fig:chandraxraylightcurves}
\end{figure*}

\section{X-ray properties of the program stars}\label{sec:Xrayproperties}

The optical and the IR analysis of Star\,1 (CXOU\,J052218.4+332820) and Star\,2 (CXOU\,J052304.2+332846) confirmed their status as a CBe and a HBe star, respectively. In this section, we analyze their X-ray light curves and spectra and derive their X-ray properties. Section \ref {subsec:chandradata} describes the data reduction procedure used for the extraction of X-ray light curves and spectra.
    
\subsection{X-ray timing analysis}\label{sec:temporalanalysis}
The multi-epoch X-ray light curves for the two target sources were extracted with an adequate time resolution and statistics in the 0.3\,--\,8 keV energy range, as mentioned in Section \ref{subsec:chandradata}. We used the CIAO task \texttt{glvary} \citep{1992ApJ...398..146Gregory} to assess the variability of the light curves. Both our program stars have a variability index in the range 0\,--\,2, indicating that there is no significant variability \citep{2012IAUS..285..402Rots}. Following this, we performed a period analysis of the X-ray light curves for both our program stars. However, due to insufficient temporal coverage of the \textit{Chandra} observations, we did not detect any statistically significant periodicity. Further, we binned the light curves to a fixed time interval and visually inspected them to avoid empty bins. The time bin used for each source's light curve is indicated in Figure \ref{fig:chandraxraylightcurves}. We then checked for hardening by calculating the hardness ratio (HR) for these sources using the equation $\mathrm{HR\,=\,\frac{H\,-\,S}{H\,+\,S}}$. We considered the energy ranges 0.3\,--\,2 keV for the soft band (S) and 2\,--\,8 keV for the hard band (H). The mean HRs for Star\,1 and Star\,2 obtained were $\sim$\,-0.6 and -0.4, respectively, indicating that neither source shows significant spectral hardening.

\begin{figure}[!ht]
    \centering
    \subfloat{\includegraphics[width=\columnwidth]{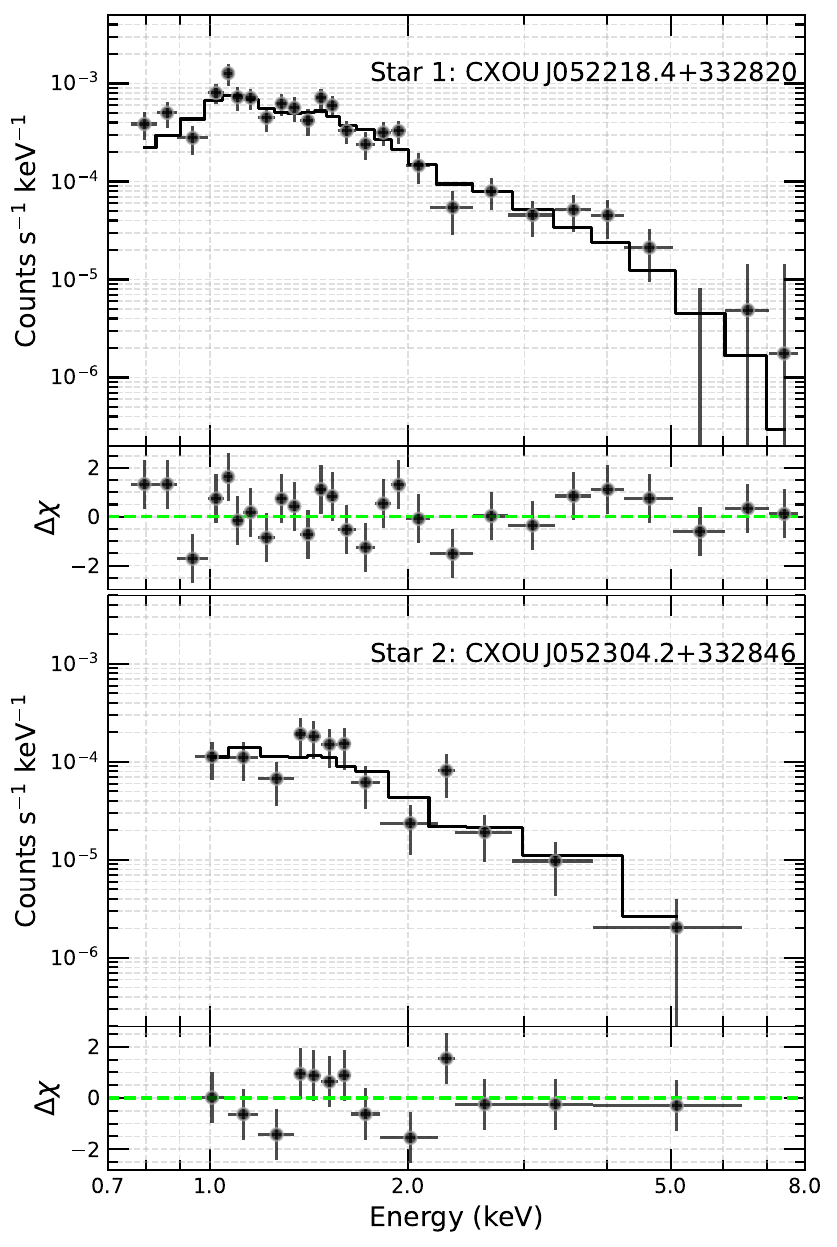}}
    \caption[A]{The \textit{Chandra} X-ray Spectrum of (top) Star\,1 (CXOU\,J052218.4+332820) and (bottom) Star\,2 (CXOU\,J052304.2+332846). Each plot shows the combined spectrum from multiple \textit{Chandra} observations. Both spectra were modeled using an absorbed single-temperature (1T) thermal plasma fit, yielding a plasma temperature kT\,$\sim$\,1.7\,--\,1.8 keV. The upper panels of both plots show the observed spectra overlaid with the best-fit model, while the lower panels present the corresponding residuals.}
    \label{fig:chandraxrayspectrum}
\end{figure}

\subsection{X-ray spectral analysis}\label{subsec:xrayspectralanalysis}
The X-ray spectra for the two sources were obtained using the CIAO script \texttt{specextract}. Each source's spectrum was grouped to ensure a minimum of 3 or more counts per bin. This grouping was determined by the quality of the spectrum, specifically the total number of spectral counts. The spectral fitting was carried out utilizing \textsc{xspec} version 12.11.1d \citep{1996ASPC..101...17Arnaud}, which is included in \textsc{heasoft} (version 6.30). The stars in our sample emit faint X-rays, so we co-added their X-ray spectra from multiple epochs, using the script \texttt{combine{\_}spectra} to obtain a combined spectrum with a better signal-to-noise ratio. This was possible as the two program stars were not variable at any of the epochs observed. We checked the variability by calculating the hardness ratio at different epochs, using the equation discussed in section \ref{sec:temporalanalysis}. We employed the plasma emission model \textsc{apec} \citep{Smith_2001} along with \textsc{tbabs} for absorption. The solar abundances were set according to \textsc{grsa} \citep{grsa1998} for the plasma model and \textsc{wilms} \citep{Wilms_2000} for the absorption model. We used a 1T \textsc{apec} model due to low spectral quality. The absorption component N$_\mathrm{H}$ of \textsc{tbabs} was fixed to values obtained from A$_\mathrm{V}$ using the equation provided by \cite{1996Ap&SS.236..285Ryter} $\mathrm{N_H}\,=\,2.0\,\times\,10^{21}\,\times\,\mathrm{A_V}$. The global metal abundances were set to 0.3 times the solar value, based on the previous X-ray studies of young stars, where sub-solar abundances provided better spectral fits \citep[refer][for further details]{2002Feigelson, 2003Imanishi, 2005Stelzer}. The plasma temperature (kT), model normalization, and X-ray flux were kept as free parameters during the spectral fitting. The combined spectra of both the program stars were fit using Cash statistics \citep[c-statistics;][]{1976A&A....52..307Cash,1979ApJ...228..939Cash}, as the data follows a Poissonian distribution (low-count). The fit quality for the c-statistic was evaluated with the \textsc{xspec} \texttt{goodness} command. To derive the unabsorbed flux values within the energy range of 0.3\,--\,8\,keV, we employed the convolution model \textsc{cflux}. The unabsorbed X-ray flux (F$_\mathrm{X}$) for the two target sources was converted to unabsorbed X-ray luminosity (L$_\mathrm{X}$) using the distances from \cite{BJ2021}. The X-ray parameters from the model fitting were determined with a 90 percent confidence interval and are tabulated in Table \ref{tab:xrayparamtable}. Figure \ref{fig:chandraxrayspectrum} shows the combined X-ray spectrum from multiple observations (see Table \ref{tab:xraydat}) for Star\,1 and Star\,2, respectively.

\begin{table*}
    \centering
    \caption{Details of X-ray parameters obtained using a single-temperature (1T) \textsc{apec} model. The model fitting was performed on the combined spectra of both program stars.}
    \footnotesize
    \setlength{\tabcolsep}{3.25pt}
    \renewcommand{\arraystretch}{1.15}

    \begin{tabular}{cccccccccc}
    \hline
        \textbf{SI No.} & \textbf{Object} & \textbf{\makecell{N$_\mathrm{H}$\\(10$^{22}$ cm$^{-2}$)}} & \textbf{\makecell{logF$_\mathrm{X}$\\(erg/s/cm$^2$)}} & \textbf{\makecell{logL$_\mathrm{X}$\\(erg/s)}} & \textbf{\makecell{kT1\\(keV)}} & \textbf{\makecell{EM\\(10$^{53}$ cm$^{-3}$)}} & \textbf{goodness} & \textbf{HR} & \textbf{log(L$_\mathrm{X}$/L$_\mathrm{bol}$)} \\ \hline \hline
        1 & CXOU\,J052218.4+332820 & 0.23 & -13.92$\pm$0.05 & 30.68$\pm$0.15 & 1.73$\pm$0.28 & 4.86$\pm$1.14 & 39\% & -0.57$\pm$0.06 & -5.61$^{+0.19}_{-0.22}$\\ 
        2 & CXOU\,J052304.2+332846 & 0.43 & -14.59$\pm$0.10 & 30.38$\pm$0.23 & 1.79$\pm$0.70 & 2.41$\pm$0.78 & 31\% & -0.47$\pm$0.11 & -5.51$^{+0.28}_{-0.34}$ \\ \hline
    \end{tabular}
    \label{tab:xrayparamtable}
\end{table*}

\subsection{Nature of X-ray emission in program stars} \label{sec:xraycontrolsamplecomparison}
\begin{figure*}[!ht]
    \centering
    \subfloat[][]{\includegraphics[width=0.31\linewidth]{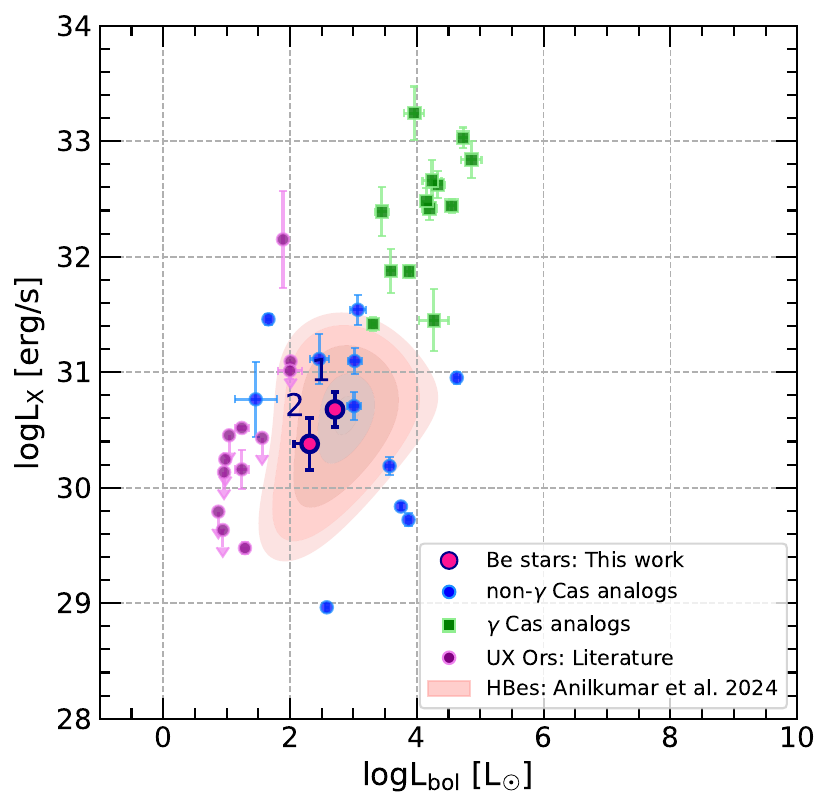}}\hspace{0.1mm}
    \subfloat[][]{\includegraphics[width=0.31\linewidth]{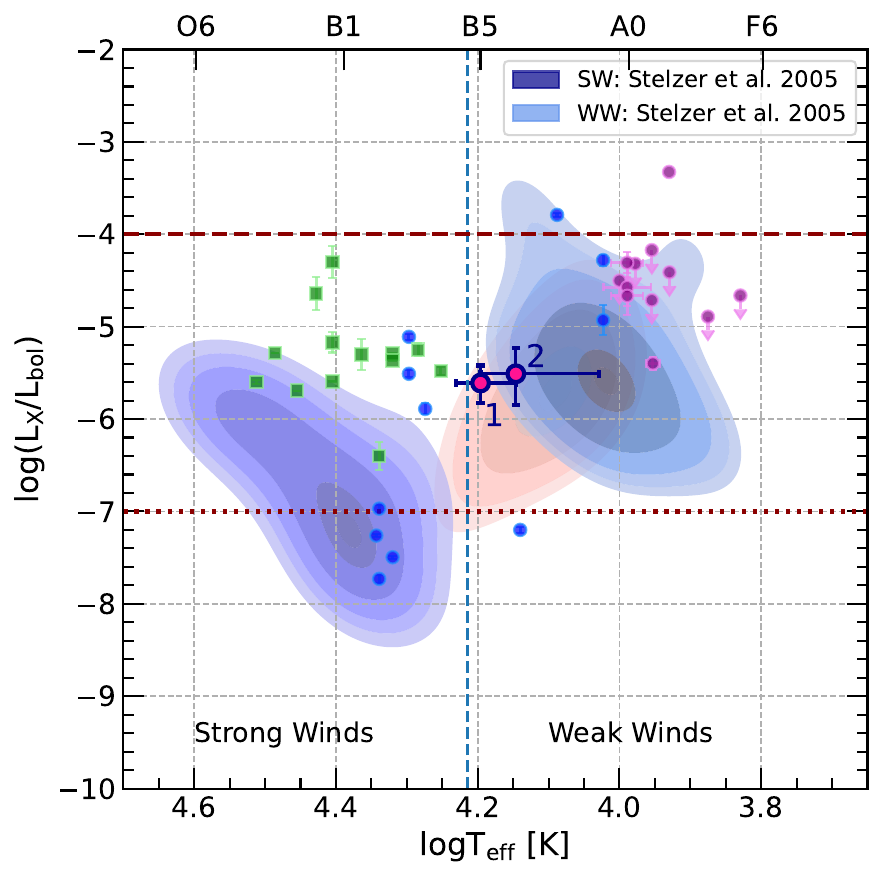}}\hspace{0.1mm}
    \subfloat[][]{\includegraphics[width=0.3\linewidth]{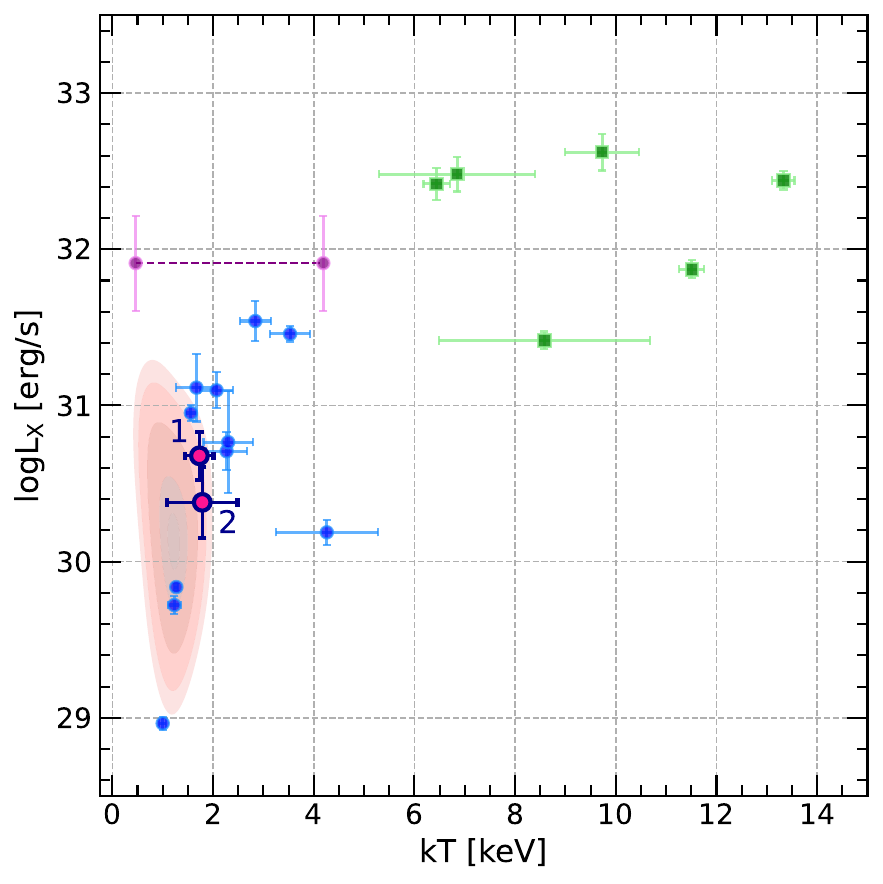}}
    \caption[A]{(a) - L$_\mathrm{X}$ versus L$_\mathrm{bol}$ plot for the two program stars (solid pink circles). Overplotted are the comparison samples of $\gamma$\,Cas stars (solid green squares), non-$\gamma$\,Cas Be stars from \cite{2018A&A...619A.148Naze} (solid blue circles), and UX\,Ori-type HAeBe stars (solid purple circles). The orange contours represent the distribution of X-ray-emitting HBe stars from \cite{2024MNRAS.530.3020Anilkumar}. (b) - L$_\mathrm{X}$/L$_\mathrm{bol}$ versus T$_\mathrm{eff}$ plot for the two program stars. Overplotted are the weak-wind and strong-wind star in blue contours described by \cite{2005Stelzer}. The red dashed line represents the canonical value of L$_\mathrm{X,P}$\,$\sim$\,10$^{-4}$\,L$_\mathrm{bol}$ for weak-wind stars, while the dotted line at L$_\mathrm{X,P}$\,$\sim$\,10$^{-7}$\,L$_\mathrm{bol}$ corresponds to strong-wind stars (see text). (c) - L$_\mathrm{X}$ versus kT for the program stars and comparison samples of HBe and CBe stars fitted with a 1T APEC model. The UX\,Ori-type HAeBe star is fitted with a 2T APEC model.}
    \label{fig:Lxvsglobparams}
\end{figure*}

From the multi-wavelength analysis in the previous section \ref{sec:analysisandresults}, it is now evident that the targets Star\,1 is a CBe star and Star\,2 is a HBe star showing UXOr-like characteristics. The main selection criterion for the program stars in our paper is that they are X-ray emitters, which is not expected for radiative stars of mid-B to late-B spectral type. In this section, we discuss X-ray parameters of the program stars, compare them with stars of their respective type from the literature, and understand the nature of the X-ray emission responsible. For this purpose, we obtained comparison samples of X-ray emitting Be stars and HAeBe stars of UX\,Ori type. The Be star sample includes $\gamma$\,Cas analogs, and non-$\gamma$\,Cas Be stars from \cite{2018A&A...619A.148Naze}. We constructed the non-$\gamma$\,Cas comparison sample from the Be stars listed in Table\,3 of \cite{2018A&A...619A.148Naze} after removing the confirmed $\gamma$\,Cas analogs identified in Table\,5 of the same study. From this sample, we further excluded sources with plasma temperatures kT\,$\sim$\,0.2\,--\,0.6\,keV and L$_{\rm X}$/L$_{\rm bol}$\,$\sim$\,10$^{-7}$\,--\,10$^{-5}$, as these objects are predominantly Oe or early-type Be stars whose X-ray emission is generally attributed to radiatively driven wind shocks \citep{Pallav1981, 1997A&A...322..167Berghoefer}. For the HAeBe comparison sample, only two stars of UX\,Ori-type were available from previous studies BF\,Ori \citep{Hamaguchi2005} and V380\,Ori \citep{2024MNRAS.530.3020Anilkumar}. We therefore searched for additional new X-ray detections of UX\,Ori-type stars from the XMM-Newton Serendipitous Source Catalog \citep[4XMM-DR14;][]{2020A&A...641A.137Traulsen} and the first eROSITA All-Sky Survey \citep[eRASS1;][]{2024A&A...682A..34Merloni}. In total, there are only 4 UX\,Ori type stars (V586\,Ori, HK\,Ori, V380\,Ori, and BF\,Ori) detected in X-rays, including the two previously reported objects. Apart from these, we find 7 UX\,Ori type stars as non-detections from the eROSITA Upper Limits catalog \citep{2024A&A...682A..35Tubinarenas,2024A&A...682A..34Merloni}. The count rates of the detected sources from these catalogs were converted from their default energy bands to unabsorbed fluxes in the 0.3\,--\,8.0\,keV band using WebPIMMS\footnote{\url{https://heasarc.gsfc.nasa.gov/cgi-bin/Tools/w3pimms/w3pimms.pl}}, assuming the same \textsc{TBABS\,$\times$\,APEC} model used in the X-ray spectral fitting (see Section \ref{subsec:xrayspectralanalysis}). The N$_{\mathrm H}$ were fixed to values obtained following the same procedure described in Section \ref{subsec:xrayspectralanalysis}, while kT was fixed at 1\,keV. The L$_{\mathrm{bol}}$, A$_{\rm V}$, and T$_{\mathrm{eff}}$ values for the comparison samples of CBe and UX\,Ori-type stars were adopted from \cite{2018A&A...619A.148Naze} and \cite{Vioque2018}, respectively. Figure \ref{fig:Lxvsglobparams} presents the X-ray properties of our program stars along with the comparison samples. 

Figure \ref{fig:Lxvsglobparams}(a) displays the L$_{\mathrm X}$--L$_{\mathrm{bol}}$ distribution of the program and comparison stars. Our program stars exhibit logL$_{\rm X(Star1)}$\,$\sim$\,30.68 and logL$_{\rm X(Star2)}$\,$\sim$\,30.38. The UX\,Ori-type stars and HBe stars from \cite{2024MNRAS.530.3020Anilkumar} span a range of L$_{\rm X}$\,$\sim$\,10$^{29.5}$\,--\,10$^{32}$\,erg\,s$^{-1}$. The non-$\gamma$\,Cas Be stars extend from $\sim$\,10$^{29}$\,--\,10$^{31.5}$\,erg\,s$^{-1}$, whereas the $\gamma$\,Cas analogs occupy the higher X-ray luminosity regime i.e., $\sim$\,10$^{31.5}$\,--\,10$^{33.5}$\,erg\,s$^{-1}$. The $\gamma$\,Cas stars also exhibit comparatively larger L$_{\rm bol}$. In contrast, the UX\,Ori-type stars, non-$\gamma$\,Cas Be stars, and HBe stars are distributed over similar ranges of both L$_{\mathrm X}$ and L$_{\mathrm{bol}}$. Both our program stars fall within this overlapping region, indicating that their X-ray luminosities are consistent with those typically observed in normal Be and HBe stars rather than the enhanced X-ray emission characteristic of $\gamma$\,Cas analogs.

Figure \ref{fig:Lxvsglobparams}(b) presents the L$_{\rm X}$/L$_{\rm bol}$--T$_{\rm eff}$ distribution of our program stars and the comparison sample. \cite{2005Stelzer}, in their study on X-ray emission from early-type stars, compared the observed X-ray luminosities (L$_{\rm X,t,c}$) of their sample with the predicted luminosities (L$_{\rm X,P}$) expected for the strong-wind (SW; SpT\,<\,B4) and the weak-wind (WW; SpT B5\,--\,A9) regimes. For SW stars, the predicted luminosity follows the canonical relation L$_{\rm X,P}$\,=\,10$^{-7}$\,L$_{\rm bol}$ \citep{1997A&A...322..167Berghoefer}, whereas for the WW stars the expected relation is L$_{\rm X,P}$\,=\,10$^{-4}$\,L$_{\rm bol}$, characteristic of low-mass magnetically active stars \citep{Preibisch2005}. \cite{2005Stelzer} further proposed that sources satisfying L$_{\rm X,t,c}$/L$_{\rm X,p}$\,<\,10 are broadly consistent with the dominant X-ray emission mechanism expected for their respective SW and WW classifications, whereas sources with L$_{\rm X,t,c}$/L$_{\rm X,p}$\,>\,10 may require an additional or alternative emission mechanism. We overplot the SW and WW samples from \cite{2005Stelzer} as contours in Figure \ref{fig:Lxvsglobparams}(b), along with canonical scaling relations at L$_{\rm X}$/L$_{\rm bol}$\,=\,10$^{-7}$ and 10$^{-4}$. Applying this criteria for our comparison sample, we find that all 4 UX\,Ori-type stars belong to the WW category and satisfy L$_{\rm X,t,c}$/L$_{\rm X,p}$\,<\,10, indicating that their X-ray properties are broadly consistent with the expected WW scaling relation. Among the 21 HBe stars, 12 belong to the WW regime and 9 to the SW regime. Of these 9 SW stars, 3 sources exhibit L$_{\rm X,t,c}$/L$_{\rm X,p}$\,>\,10. Similarly, the non-$\gamma$\,Cas Be sample consists of 7 SW and 4 WW stars, of which 3 SW stars exhibit enhanced ratios. The WW stars of all categories remain broadly consistent with the WW framework. In contrast, all 13 $\gamma$\,Cas analogs belong to the SW category, and 12 of them exhibit L$_{\rm X,t,c}$/L$_{\rm X,p}$\,>\,10. This suggests that enhanced or alternative X-ray emission mechanisms are predominantly associated with the SW population, with the most extreme deviations observed among the $\gamma$\,Cas analogs. Our program stars are located close to the transition boundary between the SW and WW regimes and exhibit logL$_{\rm X}$/L$_{\rm bol}$\,$\sim$\,-5.61 and -5.51 for Star\,1 and Star\,2, respectively. Since the program stars are classified as B5 and B7 objects, they fall within the WW regime defined by \cite{2005Stelzer}. For both our program stars, the derived ratios satisfy L$_{\rm X,t,c}$/L$_{\rm X,p}$\,<\,10, indicating the X-ray properties of our program stars are broadly consistent with the WW population exhibiting T Tauri-like magnetic X-ray characteristics. We further investigate the plasma properties of the program stars. 

Figure \ref{fig:Lxvsglobparams}(c) presents the relation between L$_{\rm X}$ and plasma temperature (kT) for the program and comparison stars. The HBe stars exhibit plasma temperatures in the range of $\sim$\,0.6\,--\,2\,keV, whereas the non-$\gamma$\,Cas Be stars show comparatively moderate plasma temperatures between $\sim$\,1\,--\,4\,keV. In contrast, the $\gamma$\,Cas analogs exhibit significantly harder X-ray emission with kT\,>\,6\,keV. Among the UX\,Ori-type stars, only V380\,Ori has a reliable estimate of plasma temperature and exhibits relatively hard X-ray emission together with flare-like variability \citep{2024MNRAS.530.3020Anilkumar}. In our previous study of X-ray-emitting HAeBe stars \citep{2024MNRAS.530.3020Anilkumar}, sources exhibiting comparatively higher plasma temperatures were often associated with flare-like variability and the presence of low-mass companions, whereas single stars generally did not exhibit strong flaring behavior across multiple epochs. Our program stars exhibit moderate plasma temperatures of kT$_{\rm (Star\,1)}$\,=\,1.73\,keV and kT$_{\rm (Star\,2)}$\,=\,1.79\,keV without any strong flares or variability across the multi-epoch \textit{Chandra} observations (see Figure \ref{fig:chandraxraylightcurves}). Such behavior has also been discussed for X-ray-emitting B stars in the Carina region, where \cite{2011ApJS..194....7Naze} reported that X-ray-bright B stars generally exhibit comparatively harder spectra (kT\,$\geq$\,1\,keV) and relatively constant X-ray luminosities across different energy bands. They suggested that the standard radiative wind-shock mechanism may become less dominant in some B stars, while intrinsic mechanisms, possibly related to magnetic confinement, could contribute to the observed X-ray emission.

\subsubsection{Possibility of Unresolved Companion-induced X-ray Emission}\label{sec:unresolvedcompanionanalysis}
X-ray emission from B-type stars, particularly later B-type stars, is often associated with unresolved low-mass companions because these stars possess weaker radiatively driven winds than O-type stars and lack the deep outer convective envelopes like low mass stars required to sustain solar-type magnetic dynamos \citep{Stelzer2006,Stelzer2009}. In Section \ref{sec:xraycontrolsamplecomparison}, we found that the X-ray properties of our program stars are broadly consistent with the WW population exhibiting PMS-like X-ray characteristics. We therefore check for the consistency of X-ray coordinates of Star\,1 and Star\,2 with Gaia astrometry and assess whether the observed X-ray emission is spatially consistent with their optical counterparts, further examining the possibility of unresolved companions contributing to the observed X-ray emission. For this, we used the X-ray centroids derived using \texttt{wavdetect}, and the optical positions were obtained from Gaia\,DR3 \citep{2023A&A...674A...1GaiaCollab}. First, we computed the positional offsets between the X-ray and Gaia coordinates for all five \textit{Chandra} observations using the \texttt{SkyCoord} module in the \texttt{astropy.coordinates} package within Python. From this, the mean positional offset between the X-ray and optical coordinates was estimated as $\sim$\,0.46\,arcsec for Star\,1. Since Star\,1 is detected at an off-axis angle of $\theta\,\sim$\,6.7\,arcmin, the \textit{Chandra} PSF at the source position becomes increasingly broadened and asymmetric. For Star\,2, the mean positional offset is smaller ($\sim$\,0.16\,arcsec), and the source lies closer to the aimpoint, i.e., $\theta\,\sim$\,2.6\,arcmin, where the \textit{Chandra} PSF remains comparatively sharper. At an estimated distance of $\sim$\,1.8\,kpc and $\sim$\,2.8\,kpc, an angular scale of 1\,arcsec corresponds to a projected separation of $\sim$\,1800\,AU and $\sim$\,2800\,AU, respectively. We do not detect any wide X-ray companion in the \textit{Chandra} ACIS images for either source within these spatial ranges. Therefore, the positional offset of $\sim$\,0.46 and 0.16\,arcsec for Star\,1 and Star\,2, respectively, between the X-ray and the optical counterpart indicates good positional agreement within uncertainties. This suggests that the X-ray emission is associated with their optical counterparts themselves, rather than originating from any unresolved companion stars. Further, we checked the Gaia RUWE values for our program stars. The RUWE values for our program stars are RUWE$_{\rm (Star\,1)}$\,=\,1.516 and RUWE$_{\rm (Star\,2)}$\,=\,1.803. A RUWE\,>\,1.4 is often considered indicative of possible unresolved companions \citep{2021A&A...649A...2Lindegren,2024A&A...688A...1CastroGinard}. However, variability and circumstellar disks can complicate the Gaia astrometric solution and artificially inflate the RUWE parameter \citep{2020MNRAS.496.1922Belokurov,2022RNAAS...6...18Fitton}. Both program stars in this work possess circumstellar disks and are likely observed at moderate inclinations, conditions which may introduce photo-center shifts affecting Gaia astrometry. \cite{2025ApJ...993..192Kalari} suggested a more conservative thresholds of RUWE\,>\,2 and \texttt{ipd\_fmp}\,>\,10, to be more appropriate for identifying multiplicity in Be stars. For our stars, the Gaia \texttt{ipd\_fmp} values are 1 and 0 for Star\,1 and Star\,2, respectively, well below the suggested cutoff. Therefore, although the Gaia astrometry indicates some degree of astrometric complexity, the current RUWE values do not provide strong evidence for unresolved companions in our program stars.

From the results discussed in the above section, we suggest that the X-ray properties of the program stars are broadly consistent with those of their respective control samples. In particular, the L$\rm_X$ and kT of Star\,1 are comparable to those observed for the non-$\gamma$\,Cas Be stars, while its $\log(\rm L_X/L_{\rm bol})$ falls within the overlapping range occupied by the WW and SW non-$\gamma$\,Cas stars as well as the $\gamma$\,Cas population. Thus, the X-ray properties of Star\,1 do not show the extremely hard and luminous characteristics associated with the $\gamma$\,Cas subclass, despite its relatively strong X-ray emission compared with that generally observed in CBe stars. The relatively strong X-ray emission of Star\,1 may therefore indicate an additional contribution beyond the standard wind-shock emission expected for CBe stars, although the present observations do not allow its physical origin to be precisely identified. In this context, the rotational modulation observed in its TESS light curve, attributed to evolving co-rotating circumstellar structures, provides a possible observational clue to the circumstellar processes associated with the X-ray emission. However, the available \textit{Chandra} observations do not show statistically significant X-ray variability, and the temporal coverage is insufficient to establish whether the optical modulation is physically connected to the X-ray emission. For Star\,2, the X-ray properties are broadly consistent with those of the X-ray-emitting HBe and UX\,Ori-type populations considered in this work. The X-ray emission from Star\,2 is consistent with the relatively small number of UX\,Ori-type objects known to emit X-rays. Its deep, quasi-periodic optical dips are characteristic of variable circumstellar obscuration, which could in principle influence the observed X-ray emission through line-of-sight absorption. However, the available X-ray spectra do not contain sufficient counts to constrain an additional circumstellar absorption component. Similarly, the absence of significant X-ray variability does not allow us to establish whether the optical variability observed in the ZTF light curves (Figure\,\ref{fig:ZTFDR24_LC}) is associated with changes in the X-ray emission or absorption. For both stars, the available observations therefore do not uniquely establish the origin of the X-ray emission. The positional consistency between the \textit{Chandra} sources and the Gaia counterparts supports an association of the X-ray emission with the optical counterparts, although an unresolved low-mass companion within the spatial resolution of \textit{Chandra} cannot be completely ruled out. The absence of significant X-ray variability, including flare-like activity, provides no direct evidence for companion-dominated emission and is broadly consistent with an intrinsic origin \citep{2024MNRAS.530.3020Anilkumar}. Further, such characteristics are compatible with magnetically influenced circumstellar or wind-shock scenarios, including magnetically confined wind-shock (MCWS) models \citep{1997B&M} and magnetically torqued disk models \citep{2002ApJ...578..951Cassinelli}. Magnetic activity has previously been suggested as a possible contributor to the X-ray emission observed in HAeBe stars \citep{Telleschi2007,2009A&A...494.1041Gunther,2013A&A...552A.142Gunther}, and magnetic fields have been detected in several HAeBe \citep{2009A&A...502..283Hubrig,2013AN....334.1093Hubrig} and CBe stars \citep{2005ASPC..337..275Neiner,2009AN....330..708Hubrig,hubrig2026magnetic}. Models proposed by \cite{2008ApJ...672.1174Li} suggest that magnetic fields of $\sim$100\,G may be sufficient to confine the winds of Be stars. Further high-angular-resolution observations, long-term spectroscopic monitoring, and high-resolution X-ray spectroscopy would be useful to better constrain the multiplicity, the location of the X-ray-emitting regions, and the physical origin of the X-ray emission in these systems. Simultaneous long-term optical and X-ray monitoring would also be valuable for investigating the possible connection between the observed optical variability and X-ray emission or absorption.

\section{Conclusion}\label{sec:conclusion}
We present a comprehensive multi-wavelength investigation of two X-ray-emitting Be stars CXOU\,J052218.4+332820 (Star\,1) and CXOU\,J052304.2+332846 (Star\,2), identified from the \textit{LAMOST Hot Emission-Line Stars Catalog}. Based on the combined optical, infrared, and X-ray diagnostics, we classify the two stars as a CBe and an HBe star exhibiting UX\,Ori-type characteristics. We outline the main findings from our study below.

\begin{enumerate}
\item We determine the spectral type of the two targets, Star\,1 and Star\,2, to be B5Ve and B7IVe, respectively. From the estimated spectral type, the A$_\mathrm{V}$ values were estimated to be 1.17 (Star\,1) and 2.16 (Star\,2).

\item The program stars are located within the FOV of the cluster NGC\,1893. From the Gaia membership analysis, we identify Star\,1 as a foreground object and a non-member of the cluster, while Star\,2 is consistent with cluster membership.

\item We find that the two program stars are in distinct evolutionary phases. Star\,1 shows no significant infrared excess and is consistent with a MS object, whereas Star\,2 exhibits infrared excess, characteristic of a PMS source. The SEDs of both stars further support these classifications.

\item The spectrum of Star\,1 exhibits strong H$\alpha$ emission (EW\,=\,-36.44\,$\pm$\,0.01\,{\AA}) along with double-peaked H$\beta$, Fe\textsc{ii}, O\textsc{i}, and Paschen emission lines. Using the double-peaked profiles, we estimate the line-forming regions within the circumstellar disk and find that H$\beta$ originates in the outer disk ($\sim$\,15\,R$_{\star}$), whereas the Fe\textsc{ii}, O\textsc{i}, and Paschen lines form closer to the star in the inner disk ($\sim$\,3\,--\,5\,R$_{\star}$). We also find evidence for disk build-up and dissipation episodes, confirming the presence of a transient decretion disk characteristic of CBe stars.

\item \textit{TESS} photometric analysis of Star\,1 reveals a rotational period of $\sim$0.355\,d, corresponding to a dominant frequency of f$_0$\,$\sim$\,2.82\,d$^{-1}$ detected across all seven sectors. The observed sector-to-sector variations in the frequencies (f$_0$\,$\sim$\,2.815\,--\,2.820\,d$^{-1}$ and 2f$_0$\,$\sim$\,5.560\,--\,5.636\,d$^{-1}$) and amplitudes (A$_1$\,$\sim$\,4.96\,--\,7.61\,mmag and A$_2$\,$\sim$\,0.50\,--\,0.92\,mmag) indicate rotational modulation associated with evolving co-rotating circumstellar structures. Combining the rotational period with the measured v$\sin\,i$, we derive an equatorial rotational velocity of v$_{\rm eq}$\,$\sim$\,436\,km\,s$^{-1}$ and an inclination angle of $i$\,$\sim$\,68$^\circ$, consistent with a CBe star viewed at a moderately high inclination.
 
\item The spectrum of Star\,2 exhibits strong asymmetric H$\alpha$ emission (EW\,=\,67.3$\pm$2.9\,{\AA}), which varies over time. We observe V/R\,>\,1 asymmetry in the double-peaked H$\beta$ and Paschen emission lines, suggesting a non-axisymmetric circumstellar disk. From the inverse P-Cygni H$\gamma$, and Fe\textsc{ii} profiles, we derive infall velocities of $\sim$\,130\,--\,150\,km\,s$^{-1}$, while the He\textsc{i} absorption lines are redshifted by $\sim$\,150\,--\,180\,km\,s$^{-1}$, providing direct evidence for ongoing accretion. From the H$\alpha$ emission, we derive a mass accretion rate of $\dot{\rm M}_{\rm acc}$\,$\sim$\,2\,$\times$\,10$^{-6}$\,M$_{\odot}$\,yr$^{-1}$, comparable to those commonly observed in actively accreting HBe and UX\,Ori-type systems.

\item The ZTF light curves of Star\,2 exhibit irregular large-amplitude dimming events ($\Delta$ZTF-g\,=\,2.945\,$\pm$\,0.023\,mag and $\Delta$ZTF-r\,=\,2.266\,$\pm$\,0.018\,mag) and quasi-periodic variability on timescales of $\sim$\,50\,--\,110\,d. We find that the LAMOST spectrum was obtained during the bright phase of Star\,2, consistent with the strong spectroscopic signatures observed. The observed color variations indicate a significant contribution from circumstellar disk material. Together, these results suggest variable circumstellar obscuration produced by non-stationary accretion-related structures, consistent with UX\,Ori-type behavior.

\item We find both our program stars emit moderately hard X-rays, with kT\,$\sim$\,1.7\,--\,1.8\,keV and logL$_{\rm X}$\,=\,30.65\,$\pm$\,0.07\,erg\,s$^{-1}$ for Star\,1 and 30.38\,$\pm$\,0.14\,erg\,s$^{-1}$ for Star\,2. The logL$_{\rm X}$/L$_{\rm bol}$\,$\sim$\,-5.6 place both stars in the WW regime, with X-ray properties associated with magnetically active coronae. We do not detect significant X-ray variability over a $\sim$\,3\,month baseline. Star\,1 exhibits X-ray characteristics typical of normal non-$\gamma$\,Cas CBe stars, whereas Star\,2 resembles HBe and UX\,Ori-type systems. From the consistency between the Gaia and \textit{Chandra} positions we suggest an intrinsic stellar origin for the X-ray emission.

\end{enumerate}

The optical and X-ray properties of our program stars are broadly consistent with magnetically influenced scenarios proposed for Be stars, such as magnetically confined wind-shock (MCWS) and magnetically torqued disk models. While our multi-wavelength analysis provides a consistent picture of the nature and circumstellar environments of both stars, the long-term evolution of their circumstellar material remains poorly constrained. Although the X-ray emission appears to be associated with the stars themselves, the mechanism responsible for producing the X-rays and the location of the X-ray-emitting plasma within the stellar or circumstellar environment remain uncertain. Resolving these issues will require coordinated optical and X-ray monitoring, spectropolarimetric measurements, and high-resolution X-ray spectroscopy with facilities such as \textit{Chandra}/HETGS and the upcoming NewAthena mission to directly probe the magnetic and circumstellar environments of these systems.

\section{Acknowledgments}
{We thank Dr. Nidhi Sabu and Dr. Shridharan B. for the valuable discussions and insights on the optical spectral analysis, CHRIST (Deemed to be University), Bangalore, India. This research has made use of data obtained from the \textit{Chandra} Data Archive provided by the \textit{Chandra} X-ray Center (CXC). We thank the \textit{Chandra} and HEASARC help desk for their assistance with X-ray data analysis and software-related queries. This work also used data and/or software provided by the High Energy Astrophysics Science Archive Research Center (HEASARC), a service of the Astrophysics Science Division at NASA/GSFC. Guoshoujing Telescope (the Large Sky Area Multi-Object Fiber Spectroscopic Telescope, LAMOST) is a National Major Scientific Project built by the Chinese Academy of Sciences. Funding for the project has been provided by the National Development and Reform Commission. LAMOST is operated and managed by the National Astronomical Observatories, Chinese Academy of Sciences. This paper includes data collected by the TESS mission, funded by NASA's Science Mission Directorate. It is also based on observations obtained with the Samuel Oschin 48-inch and the 60-inch Telescopes at Palomar Observatory as part of the Zwicky Transient Facility (ZTF) project. ZTF is supported by the National Science Foundation under Grants No. AST-1440341 and AST-2034437, along with institutional partners including Caltech, IPAC, the Oskar Klein Center at Stockholm University, the University of Maryland, UC Berkeley, University of Wisconsin-Milwaukee, University of Warwick, Ruhr University, Cornell University, Northwestern University, and Drexel University. Operations are conducted by COO, IPAC, and UW. This research has made use of the SIMBAD database and the VizieR catalogue access tool, operated at CDS, Strasbourg, France. We also used the Spanish Virtual Observatory tool VOSA, developed under the project funded by MCIN/AEI/10.13039/501100011033 through grant PID2020-112949GB-I00, and updated with support from the European Union's Horizon 2020 Research and Innovation Programme under Grant Agreement No. 776403 (EXOPLANETS-A). BM and SSK also acknowledge the financial support from CHRIST (Deemed to be University), Bangalore, through the SEED Money Projects (No: SMSS-2335, 11/2023 \& SMSS-2220, 12/202). The authors acknowledge the support provided by the Department of Science and Technology (DST) under the Fund for Improvement of S\&T Infrastructure (FIST) programme (SR/FST/PS-I/2022/208). BM and SSK acknowledge the support provided by the Science \& Engineering Research Board (SERB) project under the grant number CRG/2023/005271. We acknowledge the support of the Center for Research, CHRIST (Deemed to be University), Bangalore, India. SHE acknowledges the financial support from Mane Kancor Ingredients Pvt. Ltd., provided through CSR funding.}

\section{Data availability}
The data products generated in this study are available from the corresponding author upon reasonable request.

\bibliographystyle{raa}
\bibliography{ms2026-0383ref}

\appendix

\section{Details on the optical variability}\label{appendix:opticalvariability}
The frequencies and amplitudes of our program stars, obtained from the Lomb-Scargle periodogram analysis of the TESS and ZTF light curves, are listed in Tables \ref{tab:TESS_frequencies} and \ref{tab:ZTF_frequencies}. The TESS light curve data for Star\,1 was observed from 2021‑09‑16 to 2024‑12‑18 in sectors 43, 44, 45, 59, 71, 73, and 86. The frequencies and amplitudes (mmag) of the dominant and its first harmonic are tabulated in Table \ref{tab:TESS_frequencies}. The light curves for Star\,2 obtained from ZTF\,DR23 were observed from 2018‑03‑27 to 2024‑10‑30 in r-band and from 2018‑09‑23 to 2023‑02‑08 in g-band. The dominant frequency, its 1st harmonic, and two independent frequencies in both r- and g-bands are listed in Table \ref{tab:ZTF_frequencies}. 

\begin{table}[h!]
    \centering
    \caption{TESS light curve frequencies and amplitudes identified from the Lomb-Scargle periodogram. The values in parentheses denote 1$\sigma$ errors. The frequency ($\sigma_f$) and amplitude ($\sigma_A$) uncertainties were calculated using the equation in \cite{1999DSSN...13...28Montgomery} and \cite{2019MNRAS.487..304David}.}
    \normalsize
    \setlength{\tabcolsep}{2pt}
    \renewcommand{\arraystretch}{1.15}

    \begin{tabular}{ccccc}
    \hline
        \textbf{Sector} & \textbf{f$_0$(d$^{-1}$)} & \textbf{A$_1$ (mmag)} & \textbf{2f$_0$(d$^{-1}$)} & \textbf{A$_2$ (mmag)} \\ \hline\hline
        43 & 2.8192(11) &   5.454(9)   & 5.6344(69) &   0.503(5)   \\ 
        44 & 2.8161(11) &   4.961(8)   & 5.6364(63) &   0.535(5)   \\ 
        45 & 2.8151(11) &   5.545(9)   & 5.6182(55) &   0.656(6)   \\ 
        59 & 2.8196(10) &   5.329(8)   & 5.6354(43) &   0.744(5)   \\ 
        71 & 2.8198(11) &   6.388(11)  & 5.6357(56) &   0.707(6)   \\ 
        73 & 2.8194(11) &   5.853(10)  & 5.6276(64) &   0.589(6)   \\ 
        86 & 2.8194(10) &   7.606(13)  & 5.5598(51) &   0.917(8) \\ \hline
     \end{tabular}
    \label{tab:TESS_frequencies}
\end{table}

\begin{table}[!h]
    \centering
    \caption{ZTF lightcurve frequencies identified from the Lomb-Scargle periodogram. The values in parentheses denote 1$\sigma$ errors obtained from a Gaussian fit to each periodogram peak using the half-width at half-maximum \citep[HWHM;][]{2026JHEAp..5100529Wang}. This method is appropriate for the quasi-periodic nature observed in UX\,Ori–type stars.}
    \normalsize
    \setlength{\tabcolsep}{4pt}
    \renewcommand{\arraystretch}{1.15}

    \begin{tabular}{ccc}
    \hline
        \textbf{f (d$^{-1}$)} & \textbf{ZTF-g} & \textbf{ZTF-r} \\ \hline\hline
        f$_0$ & 0.00966(37) & 0.00970(25) \\ 
        2f$_0$ & 0.01901(15) & 0.01907(19) \\ 
        f$_1$ & 0.00879(33) & 0.00883(21) \\ 
        f$_2$ & 0.01299(14) & 0.01289(13) \\ \hline
    \end{tabular}
    \label{tab:ZTF_frequencies}
\end{table}

\end{document}